\documentclass[12pt]{article}
\usepackage{amsmath, amsthm, amssymb, amsfonts, enumerate,bm,url}
\usepackage[figuresright]{rotating}
\usepackage{lscape}
\usepackage{multirow}
\usepackage{natbib}

\begin{document}
\newcommand{\abs}[1]{\left\vert#1\right\vert}
\newcommand{\set}[1]{\left\{#1\right\}}
\newcommand{\eps}{\varepsilon}
\newcommand{\To}{\rightarrow}
\newcommand{\inv}{^{-1}}
\newcommand{\ihat}{\hat{\imath}}
\newcommand{\var}{\mbox{Var}}
\newcommand{\sd}{\mbox{SD}}
\newcommand{\cov}{\mbox{Cov}}
\newcommand{\f}{\frac}
\newcommand{\fI}[1]{\frac{1}{#1}}
\newcommand{\what}[1]{\widehat{#1}}
\newcommand{\hhat}[1]{\what{\what{#1}}}
\newcommand{\wtilde}[1]{\widetilde{#1}}
\newcommand{\bdot}{\bm{\cdot}}
\newcommand{\Th}{\theta}
\newcommand{\qmq}[1]{\quad\mbox{#1}\quad}
\newcommand{\qm}[1]{\quad\mbox{#1}}
\newcommand{\mq}[1]{\mbox{#1}\quad}
\newcommand{\tr}{\mbox{tr}}
\newcommand{\logit}{\mbox{logit}}
\newcommand{\noi}{\noindent}
\newcommand{\bni}{\bigskip\noindent}
\newcommand{\bul}{$\bullet$ }
\newcommand{\bias}{\mbox{bias}}
\newcommand{\conv}{\mbox{conv}}
\newcommand{\spn}{\mbox{span}}
\newcommand{\colspace}{\mbox{colspace}}
\newcommand{\mA}{\mathcal{A}}
\newcommand{\mF}{\mathcal{F}}
\newcommand{\mH}{\mathcal{H}}
\newcommand{\mI}{\mathcal{I}}
\newcommand{\mN}{\mathcal{N}}
\newcommand{\mR}{\mathcal{R}}
\newcommand{\mT}{\mathcal{T}}
\newcommand{\mX}{\mathcal{X}}
\newcommand{\mS}{\mathcal{S}}
\newcommand{\bbR}{\mathbb{R}}
\newcommand{\fweI}{FWE$_I$}
\newcommand{\fweII}{FWE$_{II}$}
\newcommand{\vphi}{\varphi}

\newtheorem{theorem}{Theorem}[section]
\newtheorem{corollary}{Corollary}[section]
\newtheorem{conjecture}{Conjecture}[section]
\newtheorem{proposition}{Proposition}[section]
\newtheorem{lemma}{Lemma}[section]
\newtheorem{definition}{Definition}[section]
\newtheorem{example}{Example}[section]
\newtheorem{remark}{Remark}[section]

\title{{\bf\Large Sequential Tests of Multiple Hypotheses Controlling False Discovery and Nondiscovery Rates}}

\author{\textsc{Jay Bartroff}\footnote{Department of Mathematics, University of Southern California, Los Angeles, California, USA. Email: \texttt{bartroff@usc.edu}} and \textsc{Jinlin Song}\footnote{Analysis Group, Inc., 333 South Hope Street, 27th Floor, Los Angeles, California, USA. Email: \texttt{jinlin.song@analysisgroup.com}}}  

\date{}
\maketitle

\abstract{We propose a general and flexible procedure for testing multiple hypotheses about sequential (or streaming) data that simultaneously controls both the false discovery rate (FDR) and false nondiscovery rate (FNR) under minimal assumptions about the data streams which may differ in distribution, dimension, and be dependent. All that is needed is a test statistic for each data stream that controls its conventional type I and II error probabilities, and no information or assumptions are required about the joint distribution of the statistics or data streams.  The procedure can be used with sequential, group sequential, truncated, or other sampling schemes.  The procedure is a natural extension of Benjamini and Hochberg's (1995) widely-used fixed sample size procedure to the domain of sequential data, with the added benefit of simultaneous FDR and FNR control that sequential sampling affords.  We prove the procedure's error control and give some tips for implementation in commonly encountered testing situations.} 

\section{Introduction}\label{sec:intro}
Multiple testing error metrics based on the false discovery proportion -- such as its expectation, the false discovery rate (FDR) -- are widely used in applications involving large or high dimensional data sets or when many comparisons are needed. These areas include high throughput gene and protein expression data, brain imaging, and astrophysics; \citet[][Section~1]{Muller07} give a variety of examples.  Since Benjamini and Hochberg's~\citeyearpar{Benjamini95} seminal paper introducing FDR and proving that Simes'~\citeyearpar{Simes86} earlier step-up procedure controls FDR, the topic and related problems such as Empirical Bayes have been active areas of research; see \citet{Efron01}, \citet{Efron02}, \citet{Genovese02}, \citet{Storey02}, \citet{Newton04}, \citet{Storey04}, and \citet{Cohen05}.

One characteristic of the data in some of the application areas mentioned above is that it arrives sequentially in time, or as a data stream. One such area is in certain types of clinical trials, in particular the setting discussed by \citet{Berry04} in which treatments are compared on the basis of a long list of adverse events affecting the patients during the trial; multiple endpoint clinical trials such as this are discussed in more detail in Sections~\ref{sec:ex.CT}  and \ref{sec:imp.CT}. Other areas of application involving multiple data streams include data from pharmacovigilance drug side effect databases \citep[e.g.,][]{Avery11}, testing for disease clusters over a spatial area \citep[e.g.,][]{Sonesson07} and closely related problems of industrial quality control \citep[see][]{Woodall06}, and testing for a signal in a noisy image \citep[e.g.,][]{Siegmund08}.

However, the particular needs of sequential data have largely been neglected in the FDR literature and most papers adopt Benjamini and Hochberg's \citeyearpar{Benjamini95} starting point, a set of $p$-values arising from fixed sample size hypothesis tests. Our goal is to introduce an FDR-controlling procedure with as much flexibility as the Benjamini-Hochberg (BH) procedure but tailored for sequential data, by allowing for accept/reject decisions in between sequential sampling of data streams.  In Section~\ref{sec:FDR.FNR} we introduce such a procedure which we call the  \textit{Sequential BH Procedure} that controls FDR as well as its type~II analog, the false nondiscovery rate (FNR, both defined below), under independence of data streams, and under arbitrary dependence with a small logarithmic inflation of the prescribed values; these results mirror the conditions under which \citet{Benjamini95} proved FDR control in their original  paper. We make minimal assumptions about the data streams which may differ in distribution and dimension.  The only thing the procedure needs is a test statistic for each data stream that controls the conventional type~I and II error probabilities, i.e., only marginal information about the individual test statistics is needed and no information or assumptions are required about the joint distribution of the data streams or statistics.  Likewise, there are no restrictions on the hypotheses that can be tested (i.e., any combination of simple/composite null and alternative hypotheses) provided the conventional type~I and II error probabilities can be controlled. The procedure can be used with sequential, group sequential, truncated, or other sampling schemes.  The simultaneous control of FDR and FNR is a feature of the sequential setting we consider, but if there is a restriction on the maximum sample size of a given stream, it may not be possible to achieve simultaneous FDR and FNR control since the needed error bounds \eqref{typeI}-\eqref{typeII} may not both be satisfied.  For this situation, or the one where FNR control is simply not a priority of the statistician, in Section~\ref{sec:rej} we give  a ``rejective'' version of the procedure which only stops early to reject null hypotheses and explicitly controls FDR but not necessarily FNR; see Section~\ref{sec:rej}. To aid with implementation of either of these procedures, in Section~\ref{sec:stats} we review  how to construct the component sequential tests and give closed-form expressions for the needed critical values in some commonly encountered testing situations. In Section~\ref{sec:fdrsim} we discuss a simulation study comparing the proposed procedure to its fixed sample size analog, and the paper concludes with a discussion of extensions and more suggestions for implementation.

On the one hand, our approach to deriving sequential procedures that control FDR is inspired by recent advances by \citet{Sarkar98}, \citet{Benjamini01}, and \citet{Storey04} broadening the conditions under which the BH procedure controls FDR and which give a better understanding of FDR in general. On the other hand, this work also springs from recent advances \citep{Bartroff10e,Bartroff14b,Bartroff15c} occurring recently for sequential procedures controlling familywise error rate~(FWER) and other error rates \citep{Bartroff18}.  Although the Sequential BH Procedure may appear similar to the sequential procedures of these authors controlling the FWER, the underlying principles of FDR and FWER are fundamentally much different and hence any similarity is only superficial.

A distinct but related sequential multiple testing set up has been considered in recent work by \citet{Chen17} and \citet{Javanmard18}.  In these papers, a sequence of null hypotheses and their $p$-values are observed over time and accept/reject decisions are made in a manner that controls FDR. Whereas this set up may be thought of as a single stream of experiments, each one already terminated and a terminal $p$-value computed,  it differs from the current paper which considers multiple streams of data and proposes procedures that decide when to terminate each stream individually. 

\section{Motivating Example: Multiple Endpoint Clinical Trials} \label{sec:ex.CT} There are  economic, administrative, and ethical reasons why data from clinical trials are often analyzed sequentially. Sequential (or group sequential) clinical trials with multiple \textit{endpoints}, or clinical outcomes of interest represented as hypotheses, are a special case of the general setup we will address in Section~\ref{sec:setup}. Suppose a trial concerns $K\ge 2$ endpoints, each represented by a null hypothesis~$H^{(k)}$, $k\in[K]$, which denotes $\{1,2,\ldots,K\}$ throughout.  Suppose patients are evaluated sequentially; the group sequential setting requires only minor modifications mentioned at the end of this paragraph.  Measurements or data are taken on the $n$th patient and we let  $X_{n}^{(k)}$ denote the vector of data on the $n$th patient concerning the $k$th endpoint.  Certain data points may be relevant for more than one endpoint so there may be substantial (if not complete) overlap between $X_{n}^{(k)}$ and $X_{n}^{(k')}$, say. But since the focus here is on procedures that stop early to accept or reject certain endpoints, we do not assume that $X_{n}^{(k)}$ and $X_{n}^{(k')}$ are necessarily identical. Thus, as the trial proceeds, if no endpoints are dropped we observe
\begin{equation}\label{data.trials}
\begin{array}{c}
X_1^{(1)}, \\
X_1^{(2)}, \\
X_1^{(3)}, \\
\vdots\\
X_1^{(K)}, \end{array}\qmq{then}
\begin{array}{c}
X_2^{(1)}, \\
X_2^{(2)}, \\
X_2^{(3)}, \\
\vdots\\
X_2^{(K)}, \end{array}\qmq{then}
\begin{array}{c}
X_3^{(1)}, \\
X_3^{(2)}, \\
X_3^{(3)}, \\
\vdots\\
X_3^{(K)}, \end{array}\ldots\qm{and so on.}
\end{equation}
If the patients are evaluated in groups, the only notational modification needed is that $X_{n}^{(k)}$ now denotes the vector of data on the $n$th \emph{group} concerning the $k$th endpoint.  The total number of endpoints~$K$ may be large, especially with recent advances in genetic testing  which allow biomarkers to be included as endpoints. Examples are the adverse event trials discussed by \citet{Berry04}, mentioned in Section~\ref{sec:intro}.

Another example is the randomized trial described by \citet{OBrien84} to compare two diabetes therapies, experimental and conventional, in the form of improvements in nerve function of  patients as measured through 34 different electromyographic (EMG) endpoints. In this case, $H^{(k)}$ ($k=1,\ldots,K=34$) is the null hypothesis of no difference between treatment and control in the change (baseline to evaluation) in the $k$th EMG variable, and $X_n^{(k)}$ is vector of differences in the $k$th EMG variable between for patients in the $n$th accrued group, which includes patients randomized to both the treatments. Although the trial described by  \citet{OBrien84} was fixed sample and hence a single group, a sequential version would result in data of the form \eqref{data.trials}.

In the previous example $X_{n}^{(1)}, X_{n}^{(2)}, \ldots, X_{n}^{(K)}$ are vectors of the same length (i.e., the number of patients in the $n$th group, or perhaps summary statistics thereof) but this is not a requirement of the procedures proposed below in which these can be of arbitrary size and shape and, further, may be dependent. This feature may be particularly useful in multiple endpoint clinical trials wherein data associated with different endpoints may be of different size but also is likely to be correlated since they are measurements on the same patient. For example, in clinical trials for AIDS treatments, it is common \citep[e.g.,][]{Fischl87} to have multiple endpoints of both the continuous and categorical types, like CD4 (T-cell) level, which is commonly modeled as a normal random variable, and the binary indicator of opportunistic infectious disease like pneumonia, modeled as a Bernoulli random variable. For a CD4 endpoint,  $X_{n}^{(k)}$ may be the difference, after minus before, in CD4 count for the $n$th patient, modeled as normally distributed with unknown mean and variance $\mu$ and $\sigma^2$, with associated endpoint $H^{(k)}: \mu\le 0$ of no positive treatment effect on CD4 count, versus the alternative $\mu\ge \delta$, where $\delta>0$ is some minimal meaningful treatment effect. For an opportunistic infectious disease endpoint, $X_{n}^{(k)}$ may be the indicator of the disease in the $n$th patient, modeled as a Bernoulli random variable taking the value $1$ with unknown probability $p$, with associated endpoint $H^{(k)}: p\le p_0$ where $p_0$ is some baseline rate of disease occurrence in healthy patients, versus the alternative $p\ge p_1$, where $p_1>p_0$ is an elevated occurrence rate of interest. We will show how to implement the test statistics and critical values for both of these examples in Section~\ref{sec:imp.CT}.

A feature of the Sequential BH Procedure defined in Section~\ref{sec:seqBH} is that it allows data streams to be ``dropped'' (i.e., sampling terminated) when no more information is needed to reach an accept/reject decision about the corresponding hypotheses. The need to drop certain endpoints in a multiple endpoint clinical trial while continuing others occurs frequently in practice since certain measurements are costly or invasive. An example is the well-known Women's Health Initiative \citep[WHI, see][]{Anderson04,Rossouw02}, one of the largest multiple endpoint randomized prevention studies of its kind. The WHI dropped the endpoints designed to investigate the effect of hormone replacement therapy on cardiovascular and cancer outcomes in 2002 and 2005, respectively, but continued to follow-up participants for dementia and other cognition-related endpoints. This portion of the study with the continued endpoints is known as the Women's Health Initiative Memory Study \citep[see][]{Espeland04,Shumaker98}.

\section{Control of FDR and FNR}\label{sec:FDR.FNR}

\subsection{General Notation and Setup}\label{sec:setup}
The methodology introduced below is to handle a general situation in which there are $K$ sequentially observable \emph{data streams}:
\begin{align}
&\mq{Data stream $1$}\;\;X_1^{(1)}, X_2^{(1)},\ldots\;\qm{from Experiment $1$}\nonumber\\
&\mq{Data stream $2$}\;\;X_1^{(2)}, X_2^{(2)},\ldots\;\qm{from Experiment $2$}\label{streams}\\
&\vdots \nonumber\\
&\mq{Data stream $K$}X_1^{(K)}, X_2^{(K)},\ldots\qm{from Experiment $K$.}\nonumber
\end{align} 
 In general we make no assumptions about the dimension  of the sequentially-observed data $X_n^{(k)}$, which may themselves be vectors of varying size, nor about the dependence structure of within-stream data $X_n^{(k)}, X_{n'}^{(k)}$ or between-stream data $X_n^{(k)}, X_{n'}^{(k')}$ ($k\ne k'$). Assume that for each data stream~$k\in[K]$ there is a parameter vector~$\theta^{(k)}\in\Theta^{(k)}$ governing that stream $X_1^{(k)}, X_2^{(k)},\ldots$, and it is desired to test  a null hypothesis $H^{(k)}\subseteq \Theta^{(k)}$ about $\theta^{(k)}$, versus an alternative hypothesis $G^{(k)}\subseteq \Theta^{(k)}$, which is disjoint from $H^{(k)}$. A null hypothesis $H^{(k)}$ is considered \textit{true} if $\theta^{(k)}\in H^{(k)}$, and $H^{(k)}$ is \textit{false} if $\theta^{(k)}\in G^{(k)}$.  The global parameter $\theta=(\theta^{(1)},\ldots,\theta^{(K)})$ is the concatenation of the individual parameters and is contained in the global parameter space $\Theta=\Theta^{(1)}\times\cdots\times \Theta^{(K)}$. 
The FDR and FNR are defined as
$$\mbox{FDR}=\mbox{FDR}(\theta)=E_{\theta}\left(\frac{V}{R\vee 1}\right)\qmq{and} \mbox{FNR} =\mbox{FNR}(\theta)=E_{\theta}\left(\frac{U}{S\vee 1}\right),$$
where $V$ is the number of true null hypotheses rejected, $R$ is the number of null hypotheses rejected, $U$ is the number of false null hypotheses accepted, $S$ is the number of null hypotheses accepted, and $x\vee y=\max\{x,y\}$.

For simplicity of presentation we adopt the fully sequential setting so that $n$ takes the values $1,2,\ldots$,  however other sampling schemes are possible with only minor changes to what follows and without changing our main result.  For example, the method presented here includes group sequential sampling, by either taking each $X_n^{(k)}$ in \eqref{streams} to be a group (i.e., vector) of data or, alternatively, by letting $n$ take values in a sample size set $\mN$ for which group sequential sampling with at most $g$ groups of size~$m$ corresponds to $\mN=\{m,2m,\ldots,  gm \}$.   For convenience we shall refer to the index~$n$ of the data~$X_n^{(k)}$ and test statistics as the ``sample size'' or ``time''. However, since different streams' data $X_n^{(k)}$ and $X_n^{(k')}$ may be vectors of different sizes, this value $n$ may not refer to the actual sample size in any given stream. For the same reason, $n$ may represent different ``information times'' across different streams, and does not necessarily have to coincide with the calendar time of a particular analysis.

The BH procedure requires only a valid $p$-value $p^{(k)}$ for each null hypothesis $H^{(k)}$ and, letting $p^{(k_1)}\le p^{(k_2)}\le\ldots\le p^{(k_K)}$, rejects $H^{(k_1)},\ldots,H^{(k_u)}$ where $u=\max\{s\in[K]: p^{(s)}\le s\alpha/K\}$ for a given desired FDR bound~$\alpha$, accepting all $H^{(k)}$ if the maximum doesn't exist. Playing a role analogous to the  $p$-values $p^{(k)}$, in our sequential setting we utilize a sequential test statistic $\Lambda^{(k)}_n=\Lambda^{(k)}_n(X_1^{(k)}, \ldots,X_n^{(k)})$ associated with each data stream~$k\in[K]$. What follows could have been formulated completely in terms of sequential $p$-values, making it look more like the BH procedure, however we have chosen to use arbitrary sequential test statistics~$\Lambda^{(k)}_n$ instead to maintain generality and to make the resulting procedure more user-friendly,  given the complexity and non-uniqueness of sequential $p$-values in all but the simplest cases; see \citet[][Chapters~8.4 and 9]{Jennison00}.  Nonetheless, sequential $p$-values can be used for the test statistics $\Lambda^{(k)}_n$. 
 
 For each data stream $k\in[K]$, the test statistic $\Lambda^{(k)}_n$ must satisfy certain error probabilities that only depend on its associated data stream $X_1^{(k)}, X_2^{(k)},\ldots$ and not on any other data streams in any multivariate way.  Specifically, given prescribed bounds $\alpha,\beta\in(0,1)$ on the FDR and FNR, we assume that for each test statistic $\Lambda^{(k)}_n$, $k\in[K]$, there exist critical values
  \begin{equation}\label{AB.ord}
A_1^{(k)}\le A_2^{(k)}\ldots\le A_K^{(k)}\le B_K^{(k)}\le B_{K-1}^{(k)}\le\ldots \le B_1^{(k)}\end{equation} such that
\begin{align}
P_{\theta^{(k)}}(\Lambda^{(k)}_n\ge B_s^{(k)}\;\mbox{some $n$,}\; \Lambda^{(k)}_{n'}>A_1^{(k)}\;\mbox{all $n'<n$})&\le \left(\frac{s}{K}\right) \alpha\qmq{for all}\theta^{(k)}\in H^{(k)},\label{typeI}\\
P_{\theta^{(k)}}(\Lambda^{(k)}_n\le A_s^{(k)}\;\mbox{some $n$,}\; \Lambda^{(k)}_{n'}<B_1^{(k)}\;\mbox{all $n'<n$})&\le \left(\frac{s}{K}\right)\beta  \qmq{for all}\theta^{(k)}\in G^{(k)},\label{typeII}
\end{align} for each $s\in[K]$.
These error bounds simply guarantee that $\Lambda_n^{(k)}$ has critical values allowing it to achieve conventional (i.e., not in any multiple testing sense) type~I and II error probabilities given by certain fractions of $\alpha$ and $\beta$, respectively.  In particular, \eqref{typeI} says that the sequential test on the $k$th data stream $X_1^{(k)}, X_2^{(k)},\ldots$ that samples until $\Lambda^{(k)}_n\not\in(A_1^{(k)}, B_s^{(k)})$, rejecting (resp.\ accepting) $H^{(k)}$ if $\Lambda^{(k)}_n$ crosses $B_s^{(k)}$ (resp.\ $A_1^{(k)}$) first, has type~I error probability no greater than $(s/K)\alpha$. Similarly, \eqref{typeII} says that the test which samples until $\Lambda^{(k)}_n\not\in (A_s^{(k)}, B_1^{(k)})$ has type~II error probability no greater than $(s/K)\beta$, for any $s\in[K]$. Below we will show that in many cases there are standard sequential statistics that satisfy these error bounds, and there are standard software packages that allow computation of the critical values, as well as even closed-form formulas in some cases. Given critical values satisfying \eqref{typeI}-\eqref{typeII}, the ordering \eqref{AB.ord} holds without loss of generality since otherwise $A_s^{(k)}$ could be replaced by $\wtilde{A}_s^{(k)}=\max\{A_1^{(k)},\ldots, A_s^{(k)}\}$ for which \eqref{typeII} would still hold, and similarly for $B_s^{(k)}$.  In addition, the critical values $A_s^{(k)}, B_s^{(k)}$ may also depend on the sample size~$n$ of the test statistic~$\Lambda_n^{(k)}$ being compared with them, however we omit this from the notation in order to not make it too cumbersome. This is because, although different test statistics will be ranked and compared with the critical values at different stages, the test statistics being compared always have the same current sample size~$n$ at the time of comparison (see the definition of the $\wtilde{\Lambda}_n^{(k)}$ in step~\ref{fdrsample-step} of the procedure's definition, below) making this dependence possible. That is, $\wtilde{\Lambda}_n^{(1)}, \wtilde{\Lambda}_n^{(2)},, \wtilde{\Lambda}_n^{(3)}\ldots$ will be ranked and compared, but never with any $\wtilde{\Lambda}_{n'}^{(k)}$ for $n'\ne n$.

The individual sequential test statistics $\Lambda_n^{(k)}$ form the building blocks of our Sequential BH Procedure, which we define in the next section. Like the original BH procedure which compares $p$-values, the sequential procedure involves ranking test statistics.  At the current level of generality, the sequential test statistics may be on completely different scales and so we must introduce \textit{standardizing functions} $\vphi^{(k)}$, $k\in[K]$, which are applied to the test statistics $\Lambda_n^{(k)}$ before ranking them, and the Sequential BH Procedure in the next section is defined in terms of standardized test statistics $\wtilde{\Lambda}_n^{(k)}=\vphi^{(k)}(\Lambda_n^{(k)})$. The only required property of the standardizing functions is that they are increasing functions such that $\vphi^{(k)}(A_s^{(k)})$ and $\vphi^{(k)}(B_s^{(k)})$ do not depend on $k$. For simplicity, here we take the $\vphi^{(k)}$ to be piecewise linear functions such that 
\begin{equation}\label{varphi}
\vphi^{(k)}(A_s^{(k)})=-(K-s+1)\qmq{and}\vphi^{(k)}(B_s^{(k)})=K-s+1\qmq{for all $k,s \in[K]$.}
\end{equation} That is, for $k\in[K]$ define
$$\vphi^{(k)}(x)=\begin{cases}
x-A_1^{(k)}-K,&\mbox{for}\;x\le A_1^{(k)}\\
\frac{x-A_s^{(k)}}{A_{s+1}^{(k)}-A_s^{(k)}}-(K-s+1),&\mbox{for}\;A_s^{(k)}\le x\le A_{s+1}^{(k)}\;\mbox{if}\;A_{s+1}^{(k)}>A_s^{(k)},\quad 1\le s<K\\
\frac{2(x-A_K^{(k)})}{B_K^{(k)}-A_K^{(k)}}-1,&\mbox{for}\;A_K^{(k)}\le x\le B_K^{(k)}\\
\frac{x-B_s^{(k)}}{B_{s-1}^{(k)}-B_s^{(k)}}+K-s+1,&\mbox{for}\;B_s^{(k)}\le x\le B_{s-1}^{(k)}\;\mbox{if}\;B_{s-1}^{(k)}>B_s^{(k)},\quad 1<s\le K\\
x-B_1^{(k)}+K,&\mbox{for}\;x\ge B_1^{(k)}.
\end{cases}$$ 

\subsection{The Sequential BH Procedure Controlling FDR and FNR} \label{sec:seqBH}
The Sequential BH Procedure is defined iteratively by defining its $j$th stage of sampling ($j=1,2,\ldots$), between which null hypotheses are accepted or rejected. Let $\mI_j$ denote the indices of the active null hypotheses (i.e., the null hypotheses that have not been accepted or rejected yet)  at the beginning of the $j$th stage of sampling, let $a_j$ (resp.\ $r_j$) be the number of null hypotheses that have been accepted (resp.\ rejected) at the beginning of the $j$th stage of sampling, and let $n_j$ denote the cumulative sample size of the active data streams at the end of the $j$th stage of sampling. Accordingly, set $\mI_1=[K]$, $a_1=r_1=0$, and $n_0=0$. The $j$th stage of sampling ($j=1,2,\ldots$) proceeds as follows.

\begin{enumerate}
\item\label{fdrsample-step} Sample the active data streams $\{X_n^{(k)}\}_{k\in\mI_j, \; n>n_{j-1}}$ until $n$ equals
\begin{equation}\label{fdrcont-samp}
n_j=\inf\{n>n_{j-1}: \wtilde{ \Lambda}^{(i(n,\ell))}_n\not\in(-(K-a_j-\ell+1), a_j+\ell),\qm{some $\ell\in[|\mI_j|]$}\},
\end{equation}
where $\wtilde{\Lambda}^{(k)}_n=\vphi^{(k)}(\Lambda^{(k)}_n)$ and $i(n,\ell)\in[K]$ denotes the index of the $\ell$th ordered active standardized statistic at sample size $n$.

\item  
\begin{enumerate}
\item\label{fdracc-step} If a lower boundary in \eqref{fdrcont-samp} has been crossed, that is, if 
\begin{equation}\label{stop.acc}
\wtilde{ \Lambda}^{(i(n_j,\ell))}_{n_j}\le  -(K-a_j-\ell+1)\qm{for some $\ell\in[|\mI_j|]$},
\end{equation}  then accept the $m_j\ge 1$ null hypotheses $$H^{(i(n_j,1))}, H^{(i(n_j,2))}, \ldots, H^{(i(n_j,m_j))},$$ where 
\begin{equation}\label{fdrmjacc}
m_j=\max\left\{m\le |\mI_j|: \wtilde{\Lambda}^{(i(n_j,m))}_{n_j}\le -(K-a_j-m+1)\right\},
\end{equation} and set $a_{j+1}=a_j+m_j$. Otherwise set $a_{j+1}=a_j$.

\item\label{fdrrej-step} If an upper boundary in \eqref{fdrcont-samp} has been crossed, that is, if 
\begin{equation*}
\wtilde{ \Lambda}^{(i(n_j,\ell))}_{n_j}\ge a_j+\ell\qm{for some $\ell\in[|\mI_j|]$},
\end{equation*}
 then reject the $m_j'\ge 1$ null hypotheses 
 \begin{equation}\label{fdrHsrej}
H^{(i(n_j,|\mI_j|-m_j'+1))}, H^{(i(n_j,|\mI_j|-m_j'+2))}, \ldots H^{(i(n_j,|\mI_j|))}, \end{equation}
where 
\begin{equation}\label{fdrmjrej}
m_j'=\max\left\{m\le |\mI_j|: \wtilde{\Lambda}^{(i(n_j,|\mI_j|-m+1))}_{n_j}\ge K-r_j-m+1 \right\},
\end{equation} 
and set $r_{j+1}=r_j+m_j'$. Otherwise set $r_{j+1}=r_j$.
\end{enumerate}

\item\label{fdrstop-step} Stop if there are no remaining active hypotheses, i.e., if $a_{j+1}+r_{j+1}=K$.  Otherwise, let $\mI_{j+1}$ be the indices of the remaining active hypotheses and continue on to stage~$j+1$.
\end{enumerate} 
In other words, the procedure samples all active data streams until at least one of the active null hypotheses will be accepted or rejected, indicated by the stopping rule~\eqref{fdrcont-samp}. At this point, ``step-up'' acceptance and rejection rules \eqref{fdrmjacc} and \eqref{fdrHsrej}, related to the BH procedure's rule, are used to accept or reject some active hypotheses in steps \eqref{fdracc-step} and \eqref{fdrrej-step}, respectively. After updating the list of active hypotheses, the process is repeated until no active hypotheses remain.

Before stating our main result in Theorem~\ref{thm:seqBH} that this procedure controls both FDR and FNR, we make some remarks about its definition.
\begin{enumerate}[(A)]
\item There will never be a conflict between the acceptances in Step~\eqref{fdracc-step} and the rejections in Step~\eqref{fdrrej-step}.  Suppose (toward contradiction) that at some stage~$j$ the rule in Step~\eqref{fdracc-step} said to accept $H^{(k)}$ while the rule in Step~\eqref{fdrrej-step} said to reject $H^{(k)}$. Then $k=i(n_j,\ell)$ for some $\ell\le m_j$ and $\ell\ge |\mI_j|-m_j'+1$.  The former implies
$$\wtilde{\Lambda}^{(k)}_{n_j}=\wtilde{\Lambda}^{(i(n_j,m))}_{n_j}\le \wtilde{\Lambda}^{(i(n_j,m_j))}_{n_j}\le -(K-a_j-m_j+1) <0$$ while the latter implies
$$\wtilde{\Lambda}^{(k)}_{n_j}=\wtilde{\Lambda}^{(i(n_j,m))}_{n_j}\ge \wtilde{\Lambda}^{(i(n_j,|\mI_j|-m_j'+1))}_{n_j} \ge K-r_j-m_j'+1>0,$$ a contradiction.

\item Ties in the order statistics $\wtilde{\Lambda}^{(k)}_{n}$ can be broken arbitrarily (at random, say) without affecting the error control proved in Theorem~\ref{thm:seqBH}, below.

\item As mentioned above, the critical values $A_s^{(k)}, B_s^{(k)}$ can also depend on the current sample size $n$ of the test statistic $\Lambda^{(k)}_n$ being compared to them, with only notational changes in the definition of the procedure and the properties proved below; to avoid overly cumbersome notation we have omitted this from the presentation. Standard group sequential stopping boundaries -- such as Pocock, O'Brien-Fleming, power family, and any others \citep[see][Chapters~2 and 4]{Jennison00} -- can be utilized for the individual test statistics in this way.

\end{enumerate}

Our main result, given in Theorem~\ref{thm:seqBH}, is that this procedure controls both FDR and FNR at the prescribed levels $\alpha$ and $\beta$ when the test statistics are independent, and controls them at slightly inflated values of $\alpha$ and $\beta$ under arbitrary dependence of data streams, with the inflation factor given by $\sum_{k=1}^K 1/k$, which is asymptotically equivalent to $\log K$ for large $K$.  This result generalizes the original result of \citet{Benjamini95} for their fixed-sample size procedure by building on the arguments of \citet{Benjamini01}, and is proved in the Appendix.

\begin{theorem}\label{thm:seqBH}Fix $\alpha,\beta\in(0,1)$ and suppose that \eqref{typeI}-\eqref{typeII} hold. Let $K_0$ and $K_1$ denote the number of true and false null hypotheses~$H^{(k)}$, respectively, and let $\Delta=\sum_{k=1}^K 1/k$.  Then, regardless of the dependence between the data streams, the Sequential BH Procedure defined above satisfies 
\begin{align}
\mbox{FDR}(\theta)&\le \Delta\left(\frac{K_0}{K}\right) \alpha \le\Delta\alpha\qm{and}\label{FDR.arb}\\
 \mbox{FNR}(\theta)&\le \Delta\left(\frac{K_1}{K}\right) \beta \le\Delta\beta\qmq{for all}\theta\in\Theta.\label{FNR.arb}
\end{align}Further, if the $K_0$ data streams corresponding to the true null hypotheses  are independent, then the Sequential BH Procedure  satisfies 
\begin{equation}\label{FDR.ind}
\mbox{FDR}(\theta)\le \left(\frac{K_0}{K}\right) \alpha \le\alpha\qmq{for all}\theta\in\Theta.
\end{equation} If the $K_1$ data streams corresponding to the false null hypotheses  are independent, then the Sequential BH Procedure satisfies 
\begin{equation}\label{FNR.ind}
\mbox{FNR}(\theta)\le \left(\frac{K_1}{K}\right) \beta \le\beta\qmq{for all}\theta\in\Theta.
\end{equation}

\end{theorem}

\section{A Rejective Sequential BH Procedure Controlling FDR}\label{sec:rej}

In some applications where sequential sampling is called for, the statistician is primarily concerned with stopping and rejecting a null hypothesis~$H^{(k)}$ if it appears to be false, but is content to continue sampling for a very long time if $H^{(k)}$ appears to be true. Such tests have been called ``power one tests'' (see \citet[][Chapter~5]{Mukhopadhyay09} or \citet[][Chapter~IV]{Siegmund85}). Some examples of this scenario are sequential monitoring of a process (such as manufacturing) where the null hypothesis represents the process being ``in control,'' or monitoring a drug being used in a population and the null hypothesis represents the drug being safe.  In this section we present a version of the Sequential BH Procedure with this property which is obtained from the Sequential BH Procedure in the previous section by, roughly speaking, ignoring the lower boundaries $A_s^{(k)}$ for the test statistics, plus a few other minor modifications. 

In addition to the scenarios above, this version may also be useful in applications where there is a restriction on the maximum sample size.  When this occurs it may not be possible to achieve the bounds \eqref{typeI} and \eqref{typeII}, and one alternative available to the statistician is to drop the requirement of guaranteed FNR control while still achieving guaranteed FDR control, which the procedure introduced below provides by only specifying rejections (and not acceptances) of null hypotheses. For this reason we call it the \textit{Rejective Sequential BH Procedure}. Even in the presence of a maximum sample size or truncation point, it may still be possible to achieve \eqref{typeII}, which is simply a (marginal) power condition on the $k$th component test statistic. The statistician can verify that \eqref{typeII} is possible by checking if the most stringent case, the $s=1$ case, of \eqref{typeII} holds for any values $A_1^{(k)}, B_1^{(k)}$. See the Discussion for an alternative approach of using $\beta$ as a parameter for obtaining multiple testing procedures with desirable properties.

Let the data streams $X_n^{(k)}$, test statistics $\Lambda_n^{(k)}$, and parameters $\theta^{(k)}$ and $\theta$ be as in Section~\ref{sec:setup}. Since only FDR will be explicitly controlled we only require specification of null hypotheses $H^{(k)}\subset\Theta^{(k)}$ and not alternative hypotheses $G^{(k)}$, and $H^{(k)}$ is \textit{true} if $\theta^{(k)}\in H^{(k)}$ and \textit{false} otherwise. As mentioned above, we also modify the fully sequential sampling setup of Section~\ref{sec:setup} to incorporate a maximum streamwise sample size (or ``truncation point'') $\overline{N}$ in \eqref{FDRtypeI} below since  this is most natural in the scenarios mentioned above, although what follows could be formulated without a truncation point or with sample sizes other than  $1,\ldots,\overline{N}$ by replacing statements like $n<\overline{N}$ in what follows by $n\in\mN$ for an arbitrary sample size set $\mN$, with only notational changes. Without the need for lower stopping boundaries, given a desired FDR bound $\alpha\in(0,1)$, for each test statistic $\Lambda_n^{(k)}$, $k\in[K]$, we only require the existence of ``upper'' critical values $B_K^{(k)}\le B_{K-1}^{(k)}\le\ldots\le B_1^{(k)}$ satisfying
\begin{equation}\label{FDRtypeI}
P_{\theta^{(k)}}\left(\Lambda_n^{(k)}\ge B_s^{(k)}\qmq{some}n<\overline{N}\right)\le \left(\frac{s}{K}\right) \alpha \qmq{for all}s\in[K],\quad  \theta^{(k)}\in H^{(k)}.
\end{equation} Similar to \eqref{typeI}, this is just a bound on the type~I error probability of the sequential test that stops and rejects $H^{(k)}$ at time $n<\overline{N}$ if $\Lambda_n^{(k)}\ge B_s^{(k)}$, and accepts $H^{(k)}$ otherwise. The standardizing functions $\vphi^{(k)}$ can be any increasing functions such that  $\vphi^{(k)}(B_s^{(k)})$ does not depend on $k$. Here we take
$$\vphi^{(k)}(x)=\begin{cases}
x-B_K^{(k)}+1,&\mbox{for}\;x\le B_K^{(k)}\\
\frac{x-B_s^{(k)}}{B_{s-1}^{(k)}-B_s^{(k)}}+K-s+1,&\mbox{for}\;B_s^{(k)}\le x\le B_{s-1}^{(k)}\;\mbox{if}\;B_{s-1}^{(k)}>B_s^{(k)},\quad 1<s\le K\\
x-B_1^{(k)}+K,&\mbox{for}\;x\ge B_1^{(k)},
\end{cases}$$ for all $k,s \in[K]$,  giving $\vphi^{(k)}(B_s^{(k)})=K-s+1$.

Letting $\mI_j$, $n_j$ be as in Section~\ref{sec:seqBH}, the $j$th stage ($j=0,1,\ldots$) of the Rejective Sequential BH Procedure is defined as follows.

\begin{enumerate}
\item Sample the active data streams $\{X_n^{(k)}\}_{k\in\mI_j, \; n>n_{j-1}}$ until $n$ equals
\begin{equation}\label{SeStepUpnj}
n_j=\overline{N}\wedge \inf \left\{n>n_{j-1}: \quad \wtilde{\Lambda}^{(i(n,\ell))}_n\ge \ell, \qmq{some}\ell\in [|\mI_j|]\right\},
\end{equation}
where $\wtilde{\Lambda}^{(k)}_n=\vphi^{(k)}(\Lambda^{(k)}_n)$ and $i(n,\ell)$ denotes the index of the $\ell$th ordered active standardized statistic at sample size $n$.

\item \begin{enumerate}
\item \label{step:SeStepUprej} If $n_j<\overline{N}$, then reject the null hypotheses $$H^{(i(n_j,\ell_j))},H^{(i(n_j,\ell_j+1))},\ldots,H^{(i(n_j,|\mI_j|))},$$ where 
\begin{equation}\label{SeStepUpellj.def}
\ell_j=\min\{ \ell\in [|\mI_j|]: \Lambda^{i(n_j,\ell)}_{n_j}\ge \ell \}.
\end{equation}
Set $\mI_{j+1}$ to be the indices of the remaining hypotheses and proceed to stage $j+1$.

\item Otherwise, $n=\overline{N}$ so accept all active hypotheses $H^{(k)}$, $k\in \mI_j$, and stop.
\end{enumerate}
\end{enumerate}

Like the Sequential BH Procedure, this procedure samples all active test statistics until at least one of them will be rejected, indicated by the stopping rule~\eqref{SeStepUpellj.def} which is similar to the BH rejection rule. Then a step-up rejection rule is used in Step~\ref{step:SeStepUprej}  to reject certain hypotheses before the next stage of sampling begins.  When the truncation point~$\overline{N}$ is reached, all remaining active hypotheses are accepted.   The next theorem shows that, similar to the Sequential BH PRocedure, the rejective procedure has guaranteed FDR control under independence of true hypotheses, and FDR control with a slight inflation factor under arbitrary dependence.

\begin{theorem}\label{thm:seqBH.rej} Fix $\alpha\in(0,1)$. In the above setup, suppose that there are $K_0$ true null hypotheses~$H^{(k)}$ and that \eqref{FDRtypeI} holds. Then the Rejective Sequential BH Procedure defined above satisfies \eqref{FDR.ind} if the $K_0$ data streams corresponding to the true null hypotheses  are independent, and it satisfies \eqref{FDR.arb} under arbitrary dependence between data streams.
\end{theorem}

The proof of Theorem~\ref{thm:seqBH.rej} is similar to that of Theorem~\ref{thm:seqBH} and thus is omitted.

\section{Implementation}\label{sec:stats}

In this section we discuss constructing sequential test statistics and critical values satisfying \eqref{typeI}-\eqref{typeII} (or \eqref{FDRtypeI} for the rejective version of the procedure) for individual data streams, and give some examples. Unlike many fixed-sample size settings, critical values for sequential (or group sequential) test statistics can rarely can  be written down as exact, closed form expressions. However, critical values for sequential test statistics are routinely computed to sufficient accuracy using software packages, Monte Carlo, or some form of distributional approximation, asymptotic or otherwise. In Section~\ref{sec:simple} we give closed-form expressions for the critical values $A_s^{(k)}, B_s^{(k)}$ satisfying \eqref{typeI}-\eqref{typeII} to a very close approximation, and which are based on the simple and widely-used Wald approximations for the critical values of the sequential probability ratio test (SPRT) for testing simple-vs.-simple hypotheses, which are routinely used as surrogates for more complicated testing situations by monotone likelihood ratio, least favorable distributions, and other similar considerations; see \citet[][Chapters~3.4 and 3.8]{Lehmann05} and \citet[][Chapter~II.3]{Siegmund85}. For more complicated testing situations, sequential generalized likelihood ratio statistics and their signed-root normal approximations are discussed in Section~\ref{sec:composite}. While these approaches will address many of the commonly-encountered testing situations, they do not cover every possible testing situation so we stress that the multiple testing procedure's FDR and FNR control will hold no matter the form of the hypotheses and test statistic provided \eqref{typeI}-\eqref{typeII} are satisfied, hence the critical values obtained in other ways than those discussed here may be used.

\subsection{Simple Hypotheses and Their Use as Surrogates for Certain Composite Hypotheses}\label{sec:simple}
In this section we show how to construct the test statistics~$\Lambda^{(k)}_n$ and critical values $A_s^{(k)}, B_s^{(k)}$ satisfying \eqref{typeI}-\eqref{typeII} for any data stream $k$ such that $H^{(k)}$ and $G^{(k)}$ are both simple hypotheses. This setting is of interest in practice because many more complicated composite hypotheses can be reduced to simple hypotheses. Indeed, \citet[][Section~1]{Muller07} point out that testing a battery of simple-vs.-simple hypothesis tests is the standard setup in most discussions of FDR in the literature. In this case the test statistics $\Lambda^{(k)}_n$ will be taken to be log-likelihood ratios because of their strong optimality properties of the resulting test, the SPRT; see \citet{Chernoff72}. In order to express the likelihood ratio tests in simple form, we now make the additional assumption that each data stream $X_1^{(k)},X_2^{(k)},\ldots$ constitutes independent and identically distributed data. However, we stress that this independence assumption is limited to \emph{within} each stream so that, for example, elements of $X_1^{(k)},X_2^{(k)},\ldots$ may be correlated with (or even identical to) elements of another stream $X_1^{(k')},X_2^{(k')},\ldots$.  

Formally we represent the simple null and alternative hypotheses $H^{(k)}$ and $G^{(k)}$ by the corresponding distinct density functions $h^{(k)}$ (null) and $g^{(k)}$ (alternative) with respect to some common $\sigma$-finite measure $\mu^{(k)}$.  The parameter space~$\Theta^{(k)}$ corresponding to this data stream is the set of all densities $f$ with respect to $\mu^{(k)}$, and $H^{(k)}$ is considered \textit{true} if the actual density~$f^{(k)}$ satisfies $f^{(k)}=h^{(k)}$ $\mu^{(k)}$-a.s., and is \textit{false} if $f^{(k)}=g^{(k)}$ $\mu^{(k)}$-a.s. The SPRT for testing $H^{(k)}: f^{(k)}=h^{(k)}$ vs.\ $G^{(k)}: f^{(k)}=g^{(k)}$ with type I and II error probabilities $\alpha$ and $\beta$, respectively, utilizes the simple log-likelihood ratio test statistic 
\begin{equation}\label{simpleLLR}
\Lambda^{(k)}_n=\sum_{j=1}^n \log\left(\frac{g^{(k)}(X_{j}^{(k)})}{h^{(k)}(X_{j}^{(k)})}\right)
\end{equation} and samples sequentially until $\Lambda^{(k)}_n\not\in(A, B)$, where the critical values $A, B$ satisfy
\begin{align}
P_{h^{(k)}}(\Lambda^{(k)}_n\ge B\;\mbox{some $n$,}\; \Lambda^{(k)}_{n'}>A\;\mbox{all $n'<n$})&\le \alpha\label{SPRT-typeI}\\
P_{g^{(k)}}(\Lambda^{(k)}_n\le A\;\mbox{some $n$,}\; \Lambda^{(k)}_{n'}<B\;\mbox{all $n'<n$})&\le\beta.\label{SPRT-typeII}
\end{align} The most simple and widely-used method for finding $A$ and $B$ is to use the closed-form \emph{Wald-approximations} $A=A_W(\alpha,\beta)$ and $B=B_W(\alpha,\beta)$, where 
\begin{equation}\label{myAB}
A_W(a,b) =\log\left(\frac{b}{1-a}\right)+\rho,\quad B_W(a,b)=\log\left(\frac{1-b}{a}\right)-\rho
\end{equation} for $a,b\in(0,1)$ such that $a+b\le 1$ and a fixed $\rho\ge 0$. The quantity $\rho$ is an adjustment to the boundaries to account for continuous test statistics whose excess over the boundary upon stopping may be smaller than discrete statistics. See \citet[][Section~3.3.1]{Hoel71} for a derivation of Wald's \citeyearpar{Wald47} original $\rho=0$ case and, based on Brownian motion approximations, \citet[][p.~50 and Chapter~X]{Siegmund85} derives the value $\rho=.583$ which has been used to improve the approximation for continuous random variables. With our multiple testing procedure we recommend using Siegmund's $\rho=.583$ for continuous test statistics and $\rho=0$ for discrete statistics.

Although, in general, the inequalities in \eqref{SPRT-typeI}-\eqref{SPRT-typeII} only hold approximately  when  using the Wald approximations $A=A_W(\alpha,\beta)$ and $B=B_W(\alpha,\beta)$, \citet{Hoel71} show that the actual type I and II error probabilities can only exceed $\alpha$ or $\beta$ by a negligibly small amount in the worst case, and the difference approaches $0$ for small $\alpha$ and $\beta$, which is relevant in the present multiple testing situation where we will utilize fractions of $\alpha$ and $\beta$. Next we use the Wald approximations to construct closed-form critical values $A_s^{(k)}$, $B_s^{(k)}$ satisfying \eqref{typeI}-\eqref{typeII}.  The simulations performed in Section~\ref{sec:fdrsim} show that this approximation does not lead to any exceedances of the desired FDR and FNR bounds even in the case of highly correlated data streams. Alternative approaches would be to use a software package,  Monte Carlo, or to replace \eqref{myAB} by $\log b$ and $-\log a$, respectively, for which \eqref{SPRT-typeI}-\eqref{SPRT-typeII} always hold \citep[see][]{Hoel71} and proceed similarly. The next theorem, proved in the Appendix, gives simple, closed form critical values \eqref{fdrAsBs} that can be used in lieu of these other methods to calculate the $2K$ critical values~$\{A_s^{(k)}, B_s^{(k)}\}_{s\in [K]}$ for a given data stream with simple hypotheses $H^{(k)}, G^{(k)}$ in the Sequential BH Procedure. Specifically, we show that when using \eqref{fdrAsBs}, the left-hand-sides of \eqref{typeI}-\eqref{typeII} equal the same quantities one would get using Wald's approximations with $s\alpha/K$ and $s\beta/K$ in place of $\alpha$ and $\beta$, hence the inequalities in \eqref{typeI}-\eqref{typeII} hold up to Wald's approximation.

\begin{theorem}\label{thm:simple} Fix $\alpha,\beta\in(0,1)$ such that $\alpha+\beta\le 1$. Suppose that, for a certain data stream~$k$, the associated hypotheses $H^{(k)}: f^{(k)}=h^{(k)}$ and $G^{(k)}: f^{(k)}=g^{(k)}$ are simple. For $a,b\in(0,1)$ such that $a+b\le 1$ let $\alpha_W^{(k)}(a,b)$ and $\beta_W^{(k)}(a,b)$ be the values of the probabilities on the left-hand-sides of \eqref{SPRT-typeI} and \eqref{SPRT-typeII}, respectively, when $\Lambda^{(k)}_n$ is given by \eqref{simpleLLR} and $A=A_W(a,b)$ and $B=B_W(a,b)$ are given by the Wald approximations \eqref{myAB}. For $s\in[K]$ define
\begin{equation*}
\alpha_s =\frac{\alpha(K-s\beta)}{K(K-\beta)},\quad  \beta_s =\frac{\beta(K-s\alpha)}{K(K-\alpha)}.
\end{equation*}
Finally, for $k\in[K]$ let $\alpha_{BH,s}^{(k)}$ and $\beta_{BH,s}^{(k)}$ denote the left-hand-sides of \eqref{typeI} and \eqref{typeII}, respectively, with $A_s^{(k)}$, $B_s^{(k)}$ given by
\begin{equation}\label{fdrAsBs}
A_s^{(k)} =\log\left(\frac{s\beta}{(1-\alpha_s)K}\right)+\rho,\quad B_s^{(k)}=\log\left(\frac{(1-\beta_s)K}{s\alpha}\right)-\rho.
\end{equation}
Then, for all $s\in[K]$,
\begin{gather}
s\alpha/K+\beta_s\le 1,\quad \alpha_s+s\beta/K\le 1,\label{sum<=1}\\
\alpha_{BH,s}^{(k)}=\alpha_W^{(k)}(s\alpha/K,\beta_s),\qmq{and} \beta_{BH,s}^{(k)}=\beta_W^{(k)}(\alpha_s,s\beta/K)\label{fdraH=aS}
\end{gather}
and therefore \eqref{typeI}-\eqref{typeII} hold, up to Wald's approximation, when using the critical values~\eqref{fdrAsBs}.
\end{theorem}

\subsubsection{Example: Exponential Families}\label{sec:exp.fam}

\noindent Suppose that a certain data stream $k$  is comprised of i.i.d.\ $d$-dimensional random vectors $X_1^{(k)}, X_2^{(k)},\ldots$ from a multiparameter exponential family of densities 
\begin{equation}\label{exp.fam}
X_n^{(k)}\sim f_{\theta^{(k)}}(x)=\exp[\theta^{(i)T} x-\psi^{(k)}(\theta^{(k)})],\quad n=1,2,\ldots,
\end{equation}
 where $\theta^{(k)}$ and $x$ are $d$-vectors, $(\cdot)^T$ denotes transpose, $\psi:\mathbb{R}^d\To \mathbb{R}$ is the cumulant generating function, and it is desired to test
\begin{equation}\label{exp.hyp.simp}
H^{(k)}: \theta^{(k)}=\eta \qmq{vs.} G^{(k)}: \theta^{(k)}=\gamma
\end{equation}
for given $\eta,\gamma\in\mathbb{R}^d$. Letting $S_n^{(k)}=\sum_{j=1}^nX_j^{(k)}$, the log-likelihood ratio~\eqref{simpleLLR} in this case is
\begin{equation}\label{s.v.s.LLR}
\Lambda^{(k)}_n=(\gamma-\eta)^T S_n^{(k)}-n[\psi^{(k)}(\gamma)-\psi^{(k)}(\eta)]\end{equation} 
and, by Theorem~\ref{thm:simple}, the critical values \eqref{fdrAsBs} can be used and satisfy \eqref{typeI}-\eqref{typeII} up to Wald's approximation.

As mentioned above, many more complicated testing situations reduce to this setting.  For example, to test the hypotheses $p^{(k)}\le p_0$ vs.\ $p^{(k)}\ge p_1$ about the the success probability $p^{(k)}$ of Bernoulli trials and given values $p_0<p_1$, one may wish to instead test  $H^{(k)}: p^{(k)}= p_0$ vs.\ $G^{(k)}: p^{(k)}= p_1$  by considering the worst-case error probabilities of the original hypotheses; such simplifications are of course routine in practice. For this case the exponential family~\eqref{exp.fam} and hypotheses~\eqref{exp.hyp.simp} are given by 
\begin{align}
\theta^{(k)}&=\log[p^{(k)}/(1-p^{(k)})]\label{bino.theta}\\
\psi^{(k)}(\theta^{(k)})&=\log[1+\exp(\theta^{(k)})]\label{bino.psi}\\
\eta&=\log[p_0/(1-p_0)]\label{s.v.s.eta}\\
\gamma&=\log[p_1/(1-p_1)].\label{s.v.s.gamma}
\end{align} A simulation study of the proposed procedure's performance in this setting is presented in Section~\ref{sec:fdrsims.indept}.

\subsection{Other Composite Hypotheses}\label{sec:composite}

While many composite hypotheses can be reduced to the simple-vs.-simple situation in Section~\ref{sec:simple}, the generality of Theorem~\ref{thm:seqBH} (and Theorem~\ref{thm:seqBH.rej} for the rejective version) does not require this and allows any type of hypotheses to be tested as long as the corresponding sequential statistics satisfy \eqref{typeI}-\eqref{typeII} (or \eqref{FDRtypeI} in the rejective case). In this section we discuss the more general case of how to proceed to apply Theorem~\ref{thm:seqBH} when a certain data stream $k$ is described by a multiparameter exponential family~\eqref{exp.fam} but simple hypotheses are not appropriate; Theorem~\ref{thm:seqBH.rej} and the rejective setting are discussed further below.  

Letting $\nabla$ denote the gradient, let 
\begin{equation*}
I(\theta^{(k)},\lambda^{(k)})=(\theta^{(k)}-\lambda^{(k)})^T\nabla\psi^{(k)}(\theta^{(k)})-[\psi^{(k)}(\theta^{(k)})-\psi^{(k)}(\lambda^{(k)})]
\end{equation*} denote the Kullback-Leibler information number for the distribution~\eqref{exp.fam}, and suppose it is desired to test
\begin{equation}\label{comp.hyp}
H^{(k)}: u(\theta^{(k)})\le u_0\qmq{vs.} G^{(k)}: u(\theta^{(k)})\ge u_1
\end{equation}where $u(\cdot)$ is some continuously differentiable real-valued function such that 
\begin{equation}\label{I.mono}
\mbox{for all fixed $\theta^{(k)}$, $I(\theta^{(k)},\lambda^{(k)})$ is}\left(\begin{array}{c}
\mbox{decreasing} \\
\mbox{increasing}
\end{array}\right)\mbox{in $u(\lambda^{(k)})$} \left(\begin{array}{c}
 <  \\
 > 
\end{array}\right) u(\theta^{(k)}),
\end{equation} and $u_0<u_1$ are chosen real numbers. In other words, \eqref{I.mono} says that for any $\lambda^{(k)}, \wtilde{\lambda}^{(k)}$ such that $u(\theta^{(k)})< u(\lambda^{(k)})\le u(\wtilde{\lambda}^{(k)})$, we have $I(\theta^{(k)},\lambda^{(k)})\le  I(\theta^{(k)},\wtilde{\lambda}^{(k)})$, and a similar statement with all inequalities reversed. The family of models~\eqref{exp.fam} and general form~\eqref{comp.hyp} of the hypotheses  contain a large number of situations frequently encountered in practice, including various two (or more) population comparison tests and testing problems with nuisance parameters.  For example, the sequential Student's $t$-test problem mentioned in Section~\ref{sec:ex.CT}  is a special case of this setup; details are given in the next section.

The hypotheses  \eqref{comp.hyp} can be tested with the flexible and powerful sequential generalized likelihood ratio (GLR) statistics.  Letting $$\what{\theta}_n^{(k)}=(\nabla\psi^{(k)})^{-1}\left(\frac{1}{n}\sum_{j=1}^n X_{j}^{(k)}\right)$$ denote the maximum likelihood estimate (MLE) of $\theta$ based on the data from the first $n$ observations, define
\begin{align}
\Lambda_{H,n}^{(k)}&=n\left[\inf_{\lambda:\, u(\lambda)= u_0} I(\what{\theta}_n^{(k)},\lambda)\right],\label{logGLRH}\\
\Lambda_{G,n}^{(k)}&=n\left[\inf_{\lambda:\, u(\lambda)= u_1} I(\what{\theta}_n^{(k)},\lambda)\right],\label{logGLRG}\\
\Lambda^{(k)}_n&=\begin{cases}
+\sqrt{2n\Lambda_{H,n}^{(k)}},&\mbox{if $u(\what{\theta}_n^{(k)})>u_0$ and $\Lambda_{H,n}^{(k)}\ge \Lambda_{G,n}^{(k)}$}\label{LambdaH}\\
-\sqrt{2n\Lambda_{G,n}^{(k)}},&\mbox{otherwise.}\end{cases} 
\end{align} 
The statistics \eqref{logGLRH} and \eqref{logGLRG} are the log-GLR statistics for testing against $H^{(k)}$ and against $G^{(k)}$, respectively. For finding the critical values to satisfy \eqref{typeI} and \eqref{typeII}, Monte Carlo simulation or software packages for sequential (or group sequential) sampling of Gaussian data utilizing the large-$n$ limiting distribution of the signed roots of \eqref{logGLRH}-\eqref{logGLRG} in \eqref{LambdaH} under $u(\theta^{(k)})=u_0$ and $u_1$, respectively, can be used; see \citet[][Theorem~2]{Jennison97}.

Another commonly encountered testing situation is testing the simple null hypotheses versus the composite alternative
\begin{equation}\label{comp.hyp2}
H^{(k)}: \theta^{(k)}=\theta_0^{(k)}\qmq{vs.} G^{(k)}: \theta^{(k)}\ne\theta_0^{(k)}
\end{equation} for a give value $\theta_0^{(k)}\in\mathbb{R}^d$. However, by considering true values of $\theta^{(k)}$ arbitrarily close to $\theta_0^{(k)}$ it is clear that no test of \eqref{comp.hyp2} can control the type~II error probability for all $\theta^{(k)}\in G^{(k)}$ in general, hence it may not be possible to find a test satisfying \eqref{typeII}.  If ``early stopping'' under the null hypothesis is not a priority, then the Rejective Sequential BH Procedure in Section~\ref{sec:rej} can be used with the GLR statistic \eqref{logGLRH}, as discussed above. On the other hand, in order to use the Sequential BH version that allows early stopping under the null as well, one may need to restrict $G^{(k)}$ in some way for that to be possible, for example by modifying $G^{(k)}$ to be only the $\theta^{(k)}$ such that $||\theta^{(k)}-\theta_0^{(k)}||\ge \delta$ for some smooth norm $||\cdot||$ (such as $l^2$ norm) and value $\delta>0$. This restricted form is a special case of the framework \eqref{comp.hyp} by choosing $u(\theta^{(k)})=||\theta^{(k)}-\theta_0^{(k)}||$, $u_0=0$, and $u_1=\delta$.

\subsection{Example Revisited: Multiple Endpoint Clinical Trials}\label{sec:imp.CT} 
In this section we discuss how to implement the two component hypothesis tests given as examples of endpoints in Section~\ref{sec:ex.CT} for multiple endpoint clinical trials.

The example of testing $H^{(k)}: p\le p_0$ vs.\ $G^{(k)}: p\ge p_1>p_0$ about the success probability~$p$ of i.i.d\ Bernoulli data  $X_1^{(k)}, X_2^{(k)},\ldots$, was discussed in Section~\ref{sec:exp.fam} where the monotone likelihood ratio allowed reduction to simple hypotheses and thus the closed form expressions \eqref{fdrAsBs} can be used for the critical values. A simulation study of the Sequential BH Procedure's performance on data streams of this type is presented in Section~\ref{sec:fdrsims.indept}.
 
For testing 
\begin{equation*}
H^{(k)}: \mu\le 0\qmq{vs.}G^{(k)}: \mu\ge \delta>0
\end{equation*}
about the mean~$\mu$ of i.i.d.\ normal data $X_1^{(k)}, X_2^{(k)},\ldots$ with unknown variance~$\sigma^2$, the same immediate reduction to simple hypotheses is not possible because of the nuisance parameter~$\sigma^2$, however the sequential GLR statistics in Section~\ref{sec:composite} can handle this situation. The statistics \eqref{logGLRH}-\eqref{LambdaH} are
\begin{gather}
\Lambda_{H,n}^{(k)}=(n/2)\log\left[1+\left(\frac{\overline{X}_n^{(k)}}{\what{\sigma}_n}\right)^2\right],\quad \Lambda_{G,n}^{(k)}=(n/2)\log\left[1+\left(\frac{\overline{X}_n^{(k)}-\delta}{\what{\sigma}_n}\right)^2\right],\nonumber\\
\mq{and} \Lambda_n^{(k)}=\begin{cases}
+\sqrt{2n\Lambda_{H,n}^{(k)}},&\mbox{if $\overline{X}_n^{(k)}\ge \delta/2$}\\
-\sqrt{2n\Lambda_{G,n}^{(k)}},&\mbox{otherwise,}\end{cases} \label{t.L}
\end{gather} 
where $\overline{X}_n^{(k)}$ and $\what{\sigma}_n^2$ are the usual MLEs of $\mu$ and $\sigma^2$, respectively, based on $X_1^{(k)},\ldots, X_n^{(k)}$ \citep[see][p.~106]{Bartroff06b}. In order to compute the critical values $\{A_s^{(k)}, B_s^{(k)}\}_{s\in[K]}$ satisfying \eqref{typeI}-\eqref{typeII} for  \eqref{t.L}, \citet[][Lemma~3.1]{Bartroff14b} showed that, in this case, the left-hand sides of \eqref{typeI} and \eqref{typeII} are bounded above by
\begin{equation}\label{t.upbd}
P(t_n\ge b_{n,s}\;\mbox{some $n$,}\; t_{n'}>a_{n',1}\;\mbox{all $n'<n$})\qmq{and} P(t_n\le a_{n,s}\;\mbox{some $n$,}\; t_{n'}<b_{n',1}\;\mbox{all $n'<n$}),
\end{equation} respectively, where 
\begin{gather}
a_{n,s}=-\left\{(n-1)(\exp\{(A_s^{(k)}/n)^2\}-1)\right\}^{1/2}_,\quad b_{n,s}=\left\{(n-1)(\exp\{(B_s^{(k)}/n)^2\}-1)\right\}^{1/2}_,\nonumber\\
\mq{and} t_n= \overline{Z}_n\left/ \left\{\frac{1}{n(n-1)}\sum_{i=1}^n \left(Z_i-\overline{Z}_n\right)^2\right\}^{1/2}\right.\label{t.tn}
\end{gather} 
in which $Z_1,\ldots, Z_n$ are i.i.d.\ standard normal random variables. Thus, $t_n$ has the Student's $t$ distribution with $n-1$ degrees of freedom. Using \eqref{t.upbd},  critical values satisfying \eqref{typeI} and \eqref{typeII} can be computed using recursive numerical integration, and this is the standard method used in this setting by the many sequential and group sequential software packages that exist; see \citet[][Chapter~19]{Jennison00} and \citet[][Chapter~4.3]{Bartroff13}. Alternatively, Monte Carlo can be used in which all that is needed is the generation of the i.i.d.\ standard normal random variables $Z_i$ in \eqref{t.tn}.

In some applications it may be desired to test a composite null hypothesis versus a simple alternative hypothesis, and thus control the FNR at a particular value of the unknown parameter.  For example, in the current sequential Student's $t$-test setting, suppose it is desired to test the composite null $H^{(k)}: \mu\le 0$ versus the simple alternative $G^{(k)}: (\mu,\sigma)=(\delta,\sigma_1)$, for given values $\delta>0$, $\sigma_1>0$.  By arguments similar to those in the proof of \citet[][Lemma~3.1]{Bartroff14b}, it can be shown that for the test statistic \eqref{t.L}, the error probabilities~\eqref{typeI} and \eqref{typeII} are bounded above by the terms in \eqref{t.upbd}, but with the $a_{n,s}$ and $b_{n',1}$ in the second term in \eqref{t.upbd} replaced by 
\begin{equation*}
\wtilde{a}_{n,s}=\min\left\{a_{n,s}, -\frac{\delta\sqrt{n-1}}{2\sigma_1}\right\}\qmq{and} \wtilde{b}_{n',1}=b_{n',1}-\frac{\delta\sqrt{n'-1}}{2\sigma_1},
\end{equation*} respectively. Using these formulas, the critical values can be computed using only the standard normal distribution by either recursive numerical integration or Monte Carlo, as described in the previous paragraph.

\section{Simulation Studies}\label{sec:fdrsim}
In this section we present simulation studies comparing the Sequential BH Procedure (denoted SBH throughout this section) to the fixed sample BH procedure (denoted FBH) defined in Section~\ref{sec:setup}.  We note that there are no existing sequential competitors of SBH with which to compare.  Although the traditional BH procedure could be applied to $K$ arbitrary level-$\alpha$ sequential tests performed independently of each other, the resulting procedure would not control the FNR, nor would it likely be very efficient since the stopping rules of the individual tests would not take the other data streams into account.

In Section~\ref{sec:fdrsims.indept} we compare SBH with FBH in the context of Bernoulli data streams, the setup discussed at the end of Section~\ref{sec:exp.fam}, and in Section~\ref{sec:fdrsims.dept} we consider normal data streams, and embed the streams in a multivariate normal distribution in order to simulate various between-stream correlation structures. Both  studies use the values $\alpha=.05$ and $\beta=.2$ as the prescribed FDR and FNR bounds, respectively. This same value of $\alpha$ is used for the FBH procedure and, since the resulting procedure does not guarantee FNR control at a prescribed level, in order to compare ``apples with apples'' we have varied its fixed sample size in order to make its achieved value of FNR approximately match that of the SBH procedure. For each scenario considered below we estimate FDR, FNR, their upper bounds $K_0\alpha/K$ and $K_1\beta/K$ under independence in \eqref{FDR.ind} and \eqref{FNR.ind}, respectively, the expected total sample size $EN=E(\sum_{k=1}^K N^{(k)})$ over all the data streams where $N^{(k)}$ is the total sample size of the $k$th stream, and relative savings in sample size of SBH relative to FBH using 100,000 Monte Carlo simulated batteries of $K$ sequential tests.  Finally, we note that these two simulation studies have the property that each data stream and corresponding hypothesis test has the same structure; we emphasize that this is only for the sake of getting a clear picture of the procedures' performance and this property is not required of the Sequential BH Procedure which allows arbitrary ``mixing'' of data stream distributions and types of hypotheses.

\subsection{Independent Bernoulli Data Streams}\label{sec:fdrsims.indept}
Table~\ref{tab:fdrindependent} contains the operating characteristics of SBH and FBH for testing $K$ hypotheses of the form
\begin{equation}\label{SS.Bern}
H^{(k)}: p^{(k)}\le .4\qmq{vs.}G^{(k)}: p^{(k)}\ge .6,\quad k=1,\ldots,K,
\end{equation}
about the probability $p^{(k)}$ of success in the $k$th stream of i.i.d.\ Bernoulli data which, for the sake of illustration, were generated independently of each other; a situation with between-stream dependence is considered in the next section.  Standard errors (denoted SE) are given in parentheses.  For the SBH procedure, the sequential log likelihood ratio test statistic \eqref{s.v.s.LLR}-\eqref{s.v.s.gamma} was used for each stream with the Wald approximation critical values \eqref{fdrAsBs} with $\rho=0$. For FBH, whose sample size~$N^{(k)}$ for each stream $k$ is fixed, $p$-values were computed in the standard way as $1-F_{N^{(k)},.4}(S_{N^{(k)}}^{(k)}-1)$, where $F_{n,p}(\cdot)$ is the c.d.f.\ of the Binomial distribution with $n$ trials and probability~$p$ of success, and $S_{N^{(k)}}^{(k)}$ is the sum of the $N^{(k)}$ observations from stream~$k$.  The data was generated for each data stream with $p^{(k)}=.4$ or $.6$ and the second column of Table~\ref{tab:fdrindependent} gives the number~$K_0$ of true null hypotheses, i.e., those for which $p^{(k)}=.4$, and $K_1=K-K_0$ is the number of false null hypotheses. The final column, labeled ``Savings,'' give the percent decrease in expected total sample size~$EN$ of SBH relative to FBH. Note that no standard error is given for the expected sample size of FBH because it is fixed. 

\begin{table}[htp]
\caption{Operating characteristics of sequential (SBH) and fixed-sample (FBH) BH procedures for testing the hypotheses~\eqref{SS.Bern} about the success probabilities of i.i.d.\ Bernoulli data streams.}
\begin{center}
\scalebox{0.8}{
\begin{tabular}{|c|l|ccccccc|}
\hline
$K$&$K_0$&Procedure&FDR  (SE)&$K_0\alpha/K$& FNR (SE)&$K_1\beta/K$&$EN$ (SE)&Savings\\ \hline\hline
\multirow{6}*{2}&\multirow{2}*{2} &SBH&0.0314 (0.0063)&\multirow{2}*{0.050}&0 (0)&\multirow{2}*{0}&50.8 (1.9)&\\ 
			  &	                      &FBH&0.0315 (0.0064)&                                   &0 (0) &                                    &105               &51.62\%\\\cline{2-9}
			  &\multirow{2}*{1} &SBH&0.0157 (0.0030)&\multirow{2}*{0.025}&0.0772 (0.0059)&\multirow{2}*{0.100}&61.9 (1.0)&\\ 
			  &	                      &FBH&0.0212 (0.0031)&                                    &0.0860 (0.0065)&                                     &120&48.42\%\\ \cline{2-9}
			  &0$^*$ &SBH&0 (0)& 0 &0 (0)& 0 & 273.1 (1.5)&\\\hline\hline                   
\multirow{8}*{5}
&\multirow{2}*{5} &SBH&0.0264 (0.0035)&\multirow{2}*{0.050}&0 (0)&\multirow{2}*{0}&166.5 (2.5)&\\ 
			  &	                      &FBH&0.0238 (0.0034)&                                   &0 (0) &                                    &360               &53.75\%\\\cline{2-9}
&\multirow{2}*{3} &SBH&0.0170 (0.0023)&\multirow{2}*{0.030}&0.0412 (0.0027)&\multirow{2}*{0.080}&193.7 (1.9)&\\ 
			  &	                      &FBH&0.0198 (0.0025)&                                   &0.0430 (0.0030) &                                    &370               &47.65\%\\\cline{2-9}
			  &\multirow{2}*{2}&SBH&0.0115 (0.0017)&\multirow{2}*{0.020}&0.0628 (0.0044)&\multirow{2}*{0.120}&207.2 (1.8)&\\
			  &                             &FBH&0.0188  (0.0020)&                                   &0.0629 (0.0044) &                                   &375                &44.75\%\\ \cline{2-9}
			  &0$^*$ &SBH&0 (0)& 0 &0 (0)& 0 &767.1 (1.5)&\\\hline\hline
\multirow{10}*{10}&\multirow{2}*{10} &SBH&0.0252 (0.0032)&\multirow{2}*{0.050}&0 (0)&\multirow{2}*{0}&338.0 (3.1)&\\ 
			  &	                      &FBH&0.0285 (0.0042)&                                   &0 (0) &                                    &765               &55.82\%\\\cline{2-9}
&\multirow{2}*{8} &SBH&0.0195 (0.0026)&\multirow{2}*{0.040}&0.0201 (0.0015)&\multirow{2}*{0.040}&364.5 (3.3)&\\ 
			  &	                      &FBH&0.0291 (0.0034)&                                   &0.0280 (0.0018) &                                    &760               &52.04\%\\\cline{2-9}
			  &\multirow{2}*{5} &SBH&0.0114 (0.0014)&\multirow{2}*{0.025}&0.0512 (0.0028)&\multirow{2}*{0.100}&430.3 (3.1)&\\ 
			  &	                      &FBH&0.0191 (0.0016)&                                   &0.0533 (0.0030) &                                    &770               &44.12\%\\\cline{2-9}
			  &\multirow{2}*{2}&SBH&0.0048 (0.0007)&\multirow{2}*{0.010}&0.1015 (0.0046)&\multirow{2}*{0.160}&462.1 (3.2)&\\
			  &                             &FBH&0.0085 (0.0009)&                                   &0.1037 (0.0057) &                                   &770                &39.99\%\\ \cline{2-9}
			  &0$^*$ &SBH&0 (0)& 0 &0 (0)& 0 &1541.7 (3.2)&\\\hline\hline
\multirow{10}*{20}&\multirow{2}*{20} &SBH&0.0228 (0.0023)&\multirow{2}*{0.050}&0 (0)&\multirow{2}*{0}&703.3 (4.4)&\\ 
			  &	                      &FBH&0.0206 (0.0021)&                                   &0 (0) &                                    &1650               &57.38\%\\\cline{2-9}
			  &\multirow{2}*{16} &SBH&0.0183 (0.0019)&\multirow{2}*{0.040}&0.0191 (0.0010)&\multirow{2}*{0.040}&763.5 (4.2)&\\ 
			  &	                      &FBH&0.0274 (0.0027)&                                   &0.0204 (0.0010) &                                    &1760               &56.62\%\\\cline{2-9}
			  &\multirow{2}*{10} &SBH&0.0114 (0.0010)&\multirow{2}*{0.025}&0.0493 (0.0021)&\multirow{2}*{0.100}&891.9 (5.0)&\\ 
			  &	                      &FBH&0.0208 (0.0013)&                                   &0.0544 (0.0021) &                                    &1640               &45.62\%\\\cline{2-9}
			  &\multirow{2}*{4}&SBH&0.0047 (0.0005)&\multirow{2}*{0.010}&0.0854 (0.0039)&\multirow{2}*{0.160}&964.7 (4.5)&\\
			  &                             &FBH&0.0074 (0.0007)&                                   &0.0945 (0.0040) &                                   &1700              &43.25\%\\ \cline{2-9}
			  &0$^*$ &SBH&0 (0)& 0 &0 (0)& 0 &2141.2 (4.6)&\\\hline\hline
\end{tabular}
}
\end{center}
\label{tab:fdrindependent}
\end{table}

The SBH procedure gives a sizable reduction in expected sample size relative to FBH procedure, at least roughly 40\% in all scenarios and more than 50\% savings in some. Turning our attention to FDR and FNR, note that both procedures routinely have achieved values of FDR and FNR not only less than the prescribed levels $\alpha=.05$ and $\beta=.2$, but also well below the bounds $K_0\alpha/K$ and $K_1\beta/K$, respectively. The sample size savings of SBH seems to grow with both the number~$K$ of hypotheses and the number~$K_0$ of true null hypotheses.

 An important consideration when choosing any statistical test, whether sequential or fixed sample, of hypotheses like \eqref{SS.Bern} is its performance when $p^{(k)}$ lies in the ``indifference region'' between $p_0$ and $p_1$.  Although FDR and FNR are not defined in this case, we recommend still considering other operating characteristics such as expected sample size for these values of the parameters.  For example, in the setting of Table~\ref{tab:fdrindependent}, if $p^{(k)}=.5$ for all $K=10$ streams then the expected sample size~$EN$ of the SBH procedure is 640.9, with a standard error of 2.3, based on 100,000 Monte Carlo replications. Because FDR and FNR are not defined in this case, there is no natural way to match error rates to compare with a fixed sample size procedure. However, this value of SBH's $EN$ is substantially smaller than  FBH's $EN$ even in the more favorable  scenarios (i.e., with all $p^{(k)}$ \emph{outside} the indifference region) for the latter in the $K=10$ cases of Table~\ref{tab:fdrindependent}, in which $EN=$ 760, 770, and 770, respectively.

\subsection{Correlated Normal Data Streams}\label{sec:fdrsims.dept}
Table~\ref{tab:fdrdependent} contains the operating characteristics of SBH and FBH for testing the hypotheses
\begin{equation}\label{SS.Norm}
H^{(k)}: \theta^{(k)}\le 0\qmq{vs.}G^{(k)}: \theta^{(k)}\ge\delta,\quad k=1,\ldots,K,
\end{equation}
about the mean $\theta^{(k)}$ of the $k$th stream of normal observations with variance~$1$, and where $\delta=1$.  As discussed above in Section~\ref{sec:composite}, this alternative hypothesis~$G^{(k)}$ can be thought of as a surrogate for the alternative hypothesis $\theta^{(k)}>0$. In order to generate $K$ normal data streams under various correlation structures, the $K$ streams were generated as components of a $K$-dimensional multivariate normal distribution with mean $\theta=(\theta^{(1)},\ldots,\theta^{(K)})$, given in the second column of Table~\ref{tab:fdrdependent}, and various non-identity covariance matrices $M_i$, given in the appendix.  These covariance matrices provide  a variety of different scenarios with positively and/or negatively correlated data streams. The Wald approximation critical values \eqref{fdrAsBs} were used with the continuity correction $\rho=.583$ suggested by \citet[][p.~50 and Chapter~X]{Siegmund85}. The other columns have the same meaning as in Table~\ref{tab:fdrindependent}. The $p$-values for FBH were computed in the standard way as $1-\Phi(S_{N^{(k)}}^{(k)}/\sqrt{N^{(k)}})$, where $\Phi$ is the c.d.f.\ of the standard normal distribution and $S_{N^{(k)}}^{(k)}$ is the sum of the $N^{(k)}$ observations from stream~$k$.

\begin{table}[htp]
\caption{Operating characteristics of sequential (SBH) and fixed-sample (FBH) BH procedures for testing \eqref{SS.Norm} about the means of correlated normal data streams.}
\begin{center}
\scalebox{0.7}{
\begin{tabular}{|c|c|cccccccc|}
\hline
Covariance&True $\theta$&Procedure&FDR  (SE)&$K_0\alpha/K$&$\Delta\alpha$&FNR  (SE)&$K_1 \beta/K$&$EN$ &Savings\\ \hline\hline
\multirow{2}*{$M_1$}&\multirow{2}*{$(1,0)$} &SBH&0.0249 (0.0035)&\multirow{2}*{0.025}&\multirow{2}*{0.075}&0.0983 (0.0065)&\multirow{2}*{0.100} &9.6 (0.1)&\\
			  &                                                &FBH& 0.0248 (0.0033)&&                                    &0.0970 (0.0075)&                                    &16&35.63\%\\ \hline\hline
\multirow{2}*{$M_1$}&\multirow{2}*{$(1,0)$} &SBH&0.0228 (0.0047)&\multirow{2}*{0.025}&\multirow{2}*{0.075}&0.0676 (0.0062)&\multirow{2}*{0.100} &10.5 (0.2)&\\
			  &                                                &FBH& 0.0293 (0.0043)&&                                    &0.0626 (0.0053)&                                    &20&47.50\%\\ \hline\hline
\multirow{4}*{$M_3$}&\multirow{2}*{$(1,0,1,0)$} &SBH&0.0212 (0.0030)&\multirow{2}*{0.025}&\multirow{2}*{0.104}&0.0767 (0.0045)&\multirow{2}*{0.100} &24.0 (0.2)&\\
			  &                                                &FBH& 0.0264 (0.0034)&&                                    &0.0800 (0.0051)&                                    &40&40.00\%\\ \cline{2-10}
			  &\multirow{2}*{$(1,1,0,0)$} &SBH&0.0163 (0.0036)&\multirow{2}*{0.025}&\multirow{2}*{0.104}&0.0524 (0.0053)&\multirow{2}*{0.100} &24.1 (0.4)&\\
			  &                                                &FBH& 0.0249 (0.0042)&&                                    &0.0578 (0.0053)&                                    &44&45.23\%\\ \hline\hline
\multirow{14}*{$M_4$}&\multirow{2}*{$(1,0,0,0,0,0)$} &SBH&0.0302 (0.0047)&\multirow{2}*{0.042}&\multirow{2}*{0.123}&0.0213 (0.0016)&\multirow{2}*{0.033} &31.3 (0.3)&\\
			  &                                                              &FBH&0.0379 (0.0043)&&                                   &0.0236 (0.0017)&                                    &72&56.53\%\\ \cline{2-10}
			  &\multirow{2}*{$(1,0,0,1,0,0)$} &SBH&0.0251 (0.0034)&\multirow{2}*{0.033}&\multirow{2}*{0.123}&0.0476 (0.0027)&\multirow{2}*{0.067} &34.9 (0.3)&\\
			  &                                                &FBH& 0.0324 (0.0037)&&                                    &0.0483 (0.0029)&                                    &66 &47.12\%\\ \cline{2-10}
			  &\multirow{2}*{$(1,1,0,0,0,0)$} &SBH&0.0225 (0.0038)&\multirow{2}*{0.033}&\multirow{2}*{0.123}&0.0378 (0.0034)&\multirow{2}*{0.067} &35.1 (0.5)&\\
			  &                                                &FBH& 0.0319 (0.0039)&&                                    &0.0370 (0.0036)&                                    &72&51.25\%\\ \cline{2-10}
			  &\multirow{2}*{$(1,1,1,0,0,0)$} &SBH&0.0142 (0.0032)&\multirow{2}*{0.025}&\multirow{2}*{0.123}&0.0478 (0.0044)&\multirow{2}*{0.100} &38.3 (0.6)&\\
			  &                                                &FBH& 0.0250 (0.0038)&&                                    &0.0490 (0.0048)&                                    &72&46.81\%\\ \cline{2-10}			  
			  &\multirow{2}*{$(1,1,0,1,1,0)$} &SBH&0.0137 (0.0019)&\multirow{2}*{0.017}&\multirow{2}*{0.123}&0.0952 (0.0061)&\multirow{2}*{0.133} &39.8 (0.4)&\\
			  &                                                &FBH& 0.0181 (0.0021)&&                                    &0.0879 (0.0061)&                                    &66&39.70\%\\ \cline{2-10}
			  &\multirow{2}*{$(1,1,1,1,0,0)$} &SBH&0.0113 (0.0025)&\multirow{2}*{0.017}&\multirow{2}*{0.123}&0.0826 (0.0057)&\multirow{2}*{0.133} &40.2 (0.5)&\\
			  &                                                &FBH& 0.0175 (0.0027)&&                                    &0.0884 (0.0052)&                                    &66&39.09\%\\ \cline{2-10}
			  &\multirow{2}*{$(1,1,1,1,1,0)$} &SBH&0.0069 (0.0014)&\multirow{2}*{0.008}&\multirow{2}*{0.123}&0.1174 (0.0091)&\multirow{2}*{0.167} &41.1 (0.4)&\\
			  &                                                &FBH& 0.0095 (0.0016)&&                                    &0.1226 (0.0081)&                                    &66&37.73\%\\ \hline\hline
\end{tabular}}
\end{center}
\label{tab:fdrdependent}
\end{table}

In spite of the correlations present between different data streams, the interaction of these various combinations of correlations with various true or false null hypotheses all show somewhat similar behavior to the case of independent data streams in the previous section in that SBH has sizably smaller expected sample size than FBH in all cases, roughly a 40\% reduction in most cases. Even though the independent case of Theorem~\ref{thm:seqBH} no longer applies because of the dependence, we note that in each scenario the achieved FDR and FNR rates are all less than $K_0\alpha/K$ and $K_1\beta/K$ in \eqref{FDR.ind} and \eqref{FNR.ind}, respectively.

\section{Discussion}
We have proposed a flexible procedure to combine basic sequential hypothesis tests into an FDR and FNR-controlling multiple testing procedure tailored to sequential data. The error control in Theorems~\ref{thm:seqBH} and \ref{thm:seqBH.rej} is proved under arbitrary dependence with a small logarithmic inflation~$\Delta$ of the prescribed levels $\alpha$ and $\beta$, and which may be dispensed with under independence. These were the same conditions under which \citet{Benjamini95} proved FDR control in their original paper and, as mentioned in the introduction, recent work by \citet{Benjamini01} has broadened this from independence to positive regression dependence. We fully expect to be able to similarly extend the conditions under which (uninflated) FDR control holds in the sequential domain too, but the distributional complications introduced by sequential sampling present additional challenges. Our conjecture is supported by the simulation studies in Section~\ref{sec:fdrsims.dept} and other simulation studies we have performed under strong positive dependence, in which not a single instance of achieved FDR or FNR has exceeded the uninflated levels $K_0\alpha/K$ and $K_1\alpha/K$ of the independent case. Moreover, the setting of Section~\ref{sec:fdrsims.dept} in which dependence exists between data streams but it may be impossible for the statistician to know or model \textit{a priori} is a prime example of where the proposed procedure may be useful since FDR and FNR can still be controlled by only knowing something about the \emph{marginal} distributions of the test statistics through \eqref{typeI}-\eqref{typeII}. The results of this section are encouraging that a sequential analog of Storey and Tibshirani's~\citeyearpar{Storey03} argument that the BH procedure controls FDR asymptotically as $K\To\infty$ under arbitrary dependence will hold as well.

The simultaneous control of FDR and FNR achievable by the Sequential BH Procedure is a byproduct  of the sequential setting and is analogous to the situation in classical single hypothesis testing where there exist sequential tests simultaneously controlling both type~I and II error probabilities at arbitrary levels \citep{Stein45}, a feat  which is impossible in general for fixed sample size tests \citep{Dantzig40}.  Also analogous to the classical setting, it may be that the statistician has a well-motivated value of the FDR bound~$\alpha$ in mind, but not necessarily a value of the FNR bound~$\beta$ (or the value $u_1$ in the composite alternative \eqref{comp.hyp}). In this case the rejective version of the Sequential BH Procedure in Section~\ref{sec:rej} may be used which only stops early to reject null hypotheses, i.e., when the data indicates that an alternative hypothesis is true. If substantial early stopping is also desired when null hypotheses are true, then we encourage the statistician to utilize the Sequential BH Procedure and to treat $\beta$ as a parameter that may be chosen to give a procedure with other desirable operating characteristics, such as expected total or streamwise maximum sample size.

In addition to the widely available software packages for computing group sequential critical values and the formulas \eqref{fdrAsBs} that can both be used to compute the $2K^2$ critical values~$\{A_s^{(k)}, B_s^{(k)}\}_{s,k\in[K]}$  of the individual sequential tests satisfying \eqref{typeI}-\eqref{typeII}, we have also mentioned Monte Carlo as an alternative. Although $2K^2$ critical values are needed in general, raising the specter of $2K^2$ different simulations studies, there are features of the problem making the actual number much smaller  and indicating that it is somewhat immune to the curse of dimensionality in many cases, which afflicts many problems in high-dimensional statistics.  For simplicity let us focus on the rejective version of the Sequential BH Procedure in Section~\ref{sec:rej}, however similar statements apply to the general version. In the rejective version,  the $K^2$ critical values~$\{B_s^{(k)}\}_{s,k\in[K]}$ satisfying \eqref{FDRtypeI} are needed in general. However, in settings like the simulation studies in Section~\ref{sec:fdrsim} where multiple data streams utilize test statistics of the same form, the actual number may be much smaller, e.g., $K$ if all tests are of the same form. Moreover, because of the nested nature of the error probabilities \eqref{FDRtypeI}, these $K$ values can be simulated in a \emph{single} Monte Carlo study by letting $B_s$ be the upper $(s\alpha/K)$-quantile of the simulated empirical distribution of the statistic $\max_{1\le n\le \overline{N}}\Lambda_n$.

\section*{Acknowledgements} Bartroff's work was partially supported by  grant DMS-1310127 from the National Science Foundation, grants R01-AA026368-01 and R01-GM068968 from the National Institutes of Health, and grant 1U01FD006549 from the U.S.\ Food and Drug Administration. The majority of Song's work was completed while a PhD student in the Department of Mathematics at the University of Southern California.

\section*{Appendix: Proofs and Details of Simulation Studies}
\subsection*{Proof of Theorem~\ref{thm:seqBH}} Fix $\theta\in\Theta$ and omit it from the notation.  First we prove \eqref{FDR.arb} and \eqref{FDR.ind}. Without loss of generality let $H^{(1)},\ldots,H^{(K_0)}$ denote the true null hypotheses, some $1\le K_0\le K$. For $k\in[K_0]$ and $s\in[K]$ define the events 
$$W_{k,s}=\{\wtilde{\Lambda}^{(k)}_n\ge K-s+1\;\mbox{some $ n,$}\;\wtilde{\Lambda}^{(k)}_{n'}>-K\;\mbox{all $n'<n$}\}$$ and, by \eqref{varphi} and \eqref{typeI}, we have
\begin{equation}\label{fdrPWir}
P(W_{k,s})=P(\Lambda^{(k)}_n\ge B^{(k)}_s\;\mbox{some $n$,}\;\Lambda^{(k)}_{n'}>A^{(k)}_1\;\mbox{all $n'<n$})\le s\alpha/K.
\end{equation}
For $v\in[K_0]$ and $t\in\{0,\ldots,K-K_0\}$ let 
\begin{align*}
\Omega_v&=\{\omega\subseteq[K_0]: |\omega|=v\},\\
V_{v,t}^{\omega}&=\{\mbox{$H^{(k)}$, $k\in\omega$, and $t$ false hypotheses rejected}\} \qm{for $\omega\in\Omega_v$, and}\\
V_{v,t}&=\bigcup_{\omega\in\Omega_v}V_{v,t}^{\omega}=\{\mbox{$v$ true and $t$ false hypotheses rejected}\},
\end{align*}  and note that this union is disjoint. 

We begin by showing that
\begin{equation}\label{fdrWV=ind}
P(W_{k,v+t}\cap V_{v,t}^{\omega} )\ge 1\{k\in\omega\}P(V_{v,t}^{\omega}),
\end{equation} 
which is trivial for $k\not\in\omega$.  To show that \eqref{fdrWV=ind} holds for $k \in\omega$ we will show that $V_{v,t}^{\omega}\subseteq W_{k,v+t}$ in this case. Consider any outcome in $V_{v,t}^\omega$. Since $H^{(k)}$ is rejected on this outcome, let $j$ denote the stage at which $H^{(k)}$ is rejected. By the definition of step~\ref{fdrrej-step} of the procedure, $k=i(n_j,\ell)$ for some $\ell\ge|\mI_j|-m_j'+1$, so
\begin{equation}\label{Lk.big}
\wtilde{\Lambda}_{n_j}^{(k)}=\wtilde{\Lambda}_{n_j}^{(i(n_j,\ell))} \ge \wtilde{\Lambda}_{n_j}^{(i(n_j,|\mI_j|-m_j'+1))}\ge K-(r_j+m_j')+1,
\end{equation} this last inequality by \eqref{fdrmjrej}. Since $r_j+m_j'=r_{j+1}$ and this value is no greater than the total number~$v+t$ of null hypotheses rejected on $V_{v,t}^\omega$,  \eqref{Lk.big} gives
\begin{equation}\label{Lk.big.vt}
\wtilde{\Lambda}_{n_j}^{(k)}\ge K-(v+t)+1.
\end{equation} Now suppose toward contradiction that $\wtilde{\Lambda}_{n'}^{(k)}\le -K$ for some $n'<n_j$.  Then, by \eqref{stop.acc}, $H^{(k)}$ would have been accepted at some stage $j'$ prior to $j$ since 
\begin{equation*}
\wtilde{\Lambda}_{n'}^{(k)}\le -K \le -(K-a_{j'}-\ell+1)
\end{equation*} for any possible value of $a_{j'}\ge 0$ and any $\ell\ge 1$, contradicting the assumption that $H^{(k)}$ is rejected at stage $j$. Thus it must be that $\wtilde{\Lambda}_{n'}^{(k)}> -K$ for all $n'<n_j$ and combining this with \eqref{Lk.big.vt} shows that this outcome is in $W_{k,v+t}$, finishing the proof of \eqref{fdrWV=ind}.

With \eqref{fdrWV=ind} established we now follow the argument of \citet[][Section~4]{Benjamini01} more directly with a few modifications.
\begin{multline}\label{fdrPVvs.avg}
\sum_{k=1}^{K_0}P(W_{k,v+t}\cap V_{v,t})=\sum_{k=1}^{K_0}\sum_{\omega\in\Omega_v}P(W_{k,v+t}\cap V_{v,t}^{\omega} )\ge\sum_{k=1}^{K_0}\sum_{\omega\in\Omega_v}1\{k\in\omega\}P(V_{v,t}^{\omega})\\
=\sum_{\omega\in\Omega_v}\sum_{k=1}^{K_0}1\{k\in\omega\}P(V_{v,t}^{\omega}) 
= \sum_{\omega\in\Omega_v} |\omega| P(V_{v,t}^{\omega})=v P(V_{v,t}).
\end{multline} Using this and the definition of FDR,
\begin{multline}\label{fdr.mid}
\mbox{FDR}=\sum_{t=0}^{K-K_0}\sum_{v=1}^{K_0}\frac{v}{v+t}P(V_{v,t}) \le\sum_{t=0}^{K-K_0}\sum_{v=1}^{K_0}\frac{v}{v+t}\left(\frac{1}{v}\sum_{k=1}^{K_0}P(W_{k,v+t}\cap V_{v,t})\right)\\
=\sum_{t=0}^{K-K_0}\sum_{v=1}^{K_0}\frac{1}{v+t} \sum_{k=1}^{K_0}P(W_{k,v+t}\cap V_{v,t}).
\end{multline} 
Define $U_{v,t,k}$ be the event in which, if $H^{(k)}$ is rejected, then $v-1$ other true and $t$ false null hypotheses are also rejected, so that $W_{k,v+t}\cap V_{v,t}=W_{k,v+t}\cap U_{v,t,k}$. Let $U_{s,k}=\bigcup_{v+t=s} U_{v,t,k}$ and note that, for any $k$, $U_{1,k},\ldots, U_{K,k}$ partition the sample space. Then, starting at  \eqref{fdr.mid},
\begin{equation}\label{fdr.ind0}
\mbox{FDR}\le\sum_{t=0}^{K-K_0}\sum_{v=1}^{K_0}\frac{1}{v+t} \sum_{k=1}^{K_0}P(W_{k,v+t}\cap U_{v,t,k}) =  \sum_{k=1}^{K_0}\sum_{s=1}^K \frac{1}{s}P(W_{k,s}\cap U_{s,k}).
\end{equation}
With the convention $W_{k,0}=\emptyset$, define 
$$p_{k,\ell,s}=P((W_{k,\ell} \setminus W_{k,\ell-1})\cap U_{s,k})\qmq{for}k\in[K_0], \ell\in[s], s\in[K].$$
Note that $W_{k,\ell-1}\subseteq W_{k,\ell}$, so $W_{k,s}=\bigcup_{\ell=1}^s (W_{k,\ell}\setminus W_{k,\ell-1})$ and this union is disjoint. Writing $W_{k,s}$ in this way in \eqref{fdr.ind0}, we have
\begin{multline*}
\mbox{FDR}\le \sum_{k=1}^{K_0}\sum_{s=1}^{K}\frac{1}{s}\sum_{\ell=1}^{s}p_{k,\ell,s} 
\le \sum_{k=1}^{K_0}\sum_{s=1}^{K} \sum_{\ell=1}^{s} \frac{p_{k,\ell,s}}{\ell} 
\le \sum_{k=1}^{K_0}\sum_{s=1}^{K} \sum_{\ell=1}^{K} \frac{p_{k,\ell,s}}{\ell} 
=  \sum_{k=1}^{K_0}\sum_{\ell=1}^{K}\frac{1}{\ell}\sum_{s=1}^{K}p_{k,\ell,s} \\
 =\sum_{k=1}^{K_0}\sum_{\ell=1}^{K}\frac{1}{\ell}P(W_{k,\ell}\setminus W_{k,\ell-1})
 =\sum_{k=1}^{K_0}\sum_{\ell=1}^{K}\frac{1}{\ell}[P(W_{k,\ell})-P(W_{k,\ell-1})]
 =\sum_{k=1}^{K_0}\left[\sum_{\ell=1}^{K}\frac{P(W_{k,\ell})}{\ell}-\sum_{\ell=0}^{K-1}\frac{P(W_{k,\ell})}{\ell+1}\right]\\
 =\sum_{k=1}^{K_0}\left[\sum_{\ell=1}^{K-1}\frac{P(W_{k,\ell})}{\ell(\ell+1)} +\frac{P(W_{k,K})}{K}-P(W_{k,0})\right]
 \le \sum_{k=1}^{K_0}\left[\sum_{\ell=1}^{K-1}\frac{\alpha}{K(\ell+1)}+\frac{\alpha}{K}\right]\qm{(by \eqref{fdrPWir})}\\
 =\sum_{k=1}^{K_0}\sum_{\ell=1}^{K}\frac{\alpha}{K\ell} =\Delta\left(\frac{K_0}{K}\right) \alpha.
 \end{multline*}
If data streams $k\in[K_0]$ are independent, returning to \eqref{fdr.ind0} we have
\begin{multline*}
\mbox{FDR}\le\sum_{k=1}^{K_0}\sum_{s=1}^K \frac{1}{s}P(W_{k,s})P(U_{s,k})
  \le \sum_{k=1}^{K_0}\sum_{s=1}^K \frac{1}{s}\left(\frac{s\alpha}{K}\right)P(U_{s,k})
  =\frac{\alpha}{K} \sum_{k=1}^{K_0}\sum_{s=1}^K P(U_{s,k})=\frac{\alpha}{K} \sum_{k=1}^{K_0}1\\
  =\left(\frac{K_0}{K}\right)\alpha,
\end{multline*}
where the second equality holds because $U_{1,k},\ldots,U_{K,k}$ partition the sample space.

The proof of FNR control is entirely symmetric and so is omitted here.  
\qed

\subsection*{Proof of Theorem~\ref{thm:simple}} We verify the first parts of \eqref{sum<=1} and \eqref{fdraH=aS}; the second parts are verified similarly. Using that $\beta\le1-\alpha$ and some calculus we have
$$\frac{s\alpha}{K}+\beta_s=\frac{s\alpha}{K}+\frac{\beta(K-s\alpha)}{K(K-\alpha)}\le \frac{s\alpha}{K}+\frac{(1-\alpha)(K-s\alpha)}{K(K-\alpha)}=\frac{s}{K}-\int_\alpha^1 \frac{(K-1)(s-1)}{(K-a)^2}da\le \frac{s}{K}\le 1.$$
The forms of $A_s^{(k)}$ and $B_s^{(k)}$ in \eqref{fdrAsBs} can equivalently be written as $A_W(\alpha_s,s\beta/K)$ and $B_W(s\alpha/K,\beta_s)$, respectively, and it is simple algebra to check that $A_W(s\alpha/K,\beta_s)=A_1^{(k)}$ for all $s\in[K]$. Then 
\begin{align*}
\alpha_{BH,s}^{(k)}&=P_{h^{(k)}}(\Lambda^{(k)}_n\ge B_s^{(k)}\mbox{ some $n$, }\Lambda^{(k)}_{n'}>A_1^{(k)}\;\mbox{all $n'<n$})\\
&=P_{h^{(k)}}(\Lambda^{(k)}_n\ge B_W(s\alpha/K,\beta_s)\;\mbox{some $n$,}\; \Lambda^{(k)}_{n'}>A_W(s\alpha/K,\beta_s)\;\mbox{all $n'<n$}) \\
&=\alpha_W^{(k)}(s\alpha/K,\beta_s),
\end{align*}
by definition of $\alpha_W^{(k)}$.\qed

\subsection*{Details of Simulation Studies}
The four covariance matrices used in the simulations in Section~\ref{sec:fdrsims.dept} are as follows:
\begin{align*}
M_1&=\left(\begin{array}{cc}
1&0.8\\
0.8&1\end{array}\right)\\
M_2&=\left(\begin{array}{cc}
1&-0.8\\
-0.8&1\end{array}\right)\\
M_3&=\left(\begin{array}{cccc}
1&0.8&-0.6&-0.8\\
0.8&1&-0.6&-0.8\\
-0.6&-0.6&1&0.8\\
-0.8&-0.8&0.8&1\end{array}\right)\\
M_4&=\left(\begin{array}{cccccc}
1&0.8&0.6&-0.4&-0.6&-0.8\\
0.8&1&0.8&-0.4&-0.6&-0.8\\
0.6&0.8&1&-0.4&-0.6&-0.8\\
-0.4&-0.4&-0.4&1&0.8&0.6\\
-0.6&-0.6&-0.6&0.8&1&0.8\\
-0.8&-0.8&-0.8&0.6&0.8&1\end{array}\right)
\end{align*}


\begin{thebibliography}{}

\bibitem[Anderson et~al., 2004]{Anderson04}
Anderson, G.~L., Limacher, M.~C., Assaf, A.~R., Bassford, T., Beresford, S.~A.,
  Black, H.~R., Bonds, D.~E., Brunner, R.~L., Brzyski, R.~G., Caan, B., et~al.
  (2004).
\newblock Effects of conjugated equine estrogen in postmenopausal women with
  hysterectomy: The {W}omen's {H}ealth {I}nitiative randomized controlled
  trial.
\newblock {\em Journal of the American Medical Association},
  291(14):1701--1712.

\bibitem[Avery et~al., 2011]{Avery11}
Avery, A.~J., Anderson, C., Bond, C., Fortnum, H., Gifford, A., Hannaford,
  P.~C., Hazell, L., Krska, J., Lee, A., Mclernon, D.~J., et~al. (2011).
\newblock Evaluation of patient reporting of adverse drug reactions to the {UK}
  `{Y}ellow {C}ard {S}cheme': literature review, descriptive and qualitative
  analyses, and questionnaire surveys.
\newblock {\em Health Technology Assessment}, 15:iii--227.

\bibitem[Bartroff, 2006]{Bartroff06b}
Bartroff, J. (2006).
\newblock Efficient three-stage {$t$}-tests.
\newblock In {\em Recent Developments in Nonparametric Inference and
  Probability: Festschrift for {M}ichael {W}oodroofe}, volume~50 of {\em IMS
  Lecture Notes Monograph Series}, pages 105--111, Hayward. Institute of
  Mathematical Statistics.

\bibitem[Bartroff, 2018]{Bartroff18}
Bartroff, J. (2018).
\newblock Multiple hypothesis tests controlling generalized error rates for
  sequential data.
\newblock {\em Statistica Sinica}, 28:363--398.

\bibitem[Bartroff and Lai, 2010]{Bartroff10e}
Bartroff, J. and Lai, T.~L. (2010).
\newblock Multistage tests of multiple hypotheses.
\newblock {\em Communications in Statistics -- Theory and Methods (Special
  Issue Honoring M. Akahira, M. Aoshima, ed.)}, 39:1597--1607.

\bibitem[Bartroff et~al., 2013]{Bartroff13}
Bartroff, J., Lai, T.~L., and Shih, M. (2013).
\newblock {\em Sequential Experimentation in Clinical Trials: Design and
  Analysis}.
\newblock Springer, New York.

\bibitem[Bartroff and Song, 2014]{Bartroff14b}
Bartroff, J. and Song, J. (2014).
\newblock Sequential tests of multiple hypotheses controlling type {I} and {II}
  familywise error rates.
\newblock {\em Journal of Statistical Planning and Inference}, 153:100--114.

\bibitem[Bartroff and Song, 2015]{Bartroff15c}
Bartroff, J. and Song, J. (2015).
\newblock A rejection principle for sequential tests of multiple hypotheses
  controlling familywise error rates.
\newblock {\em Scandinavian Journal of Statistics}, 43:3--19.

\bibitem[Benjamini and Hochberg, 1995]{Benjamini95}
Benjamini, Y. and Hochberg, Y. (1995).
\newblock Controlling the false discovery rate: {A} practical and powerful
  approach to multiple testing.
\newblock {\em Journal of the Royal Statistical Society, Series B:
  Methodological}, 57:289--300.

\bibitem[Benjamini and Yekutieli, 2001]{Benjamini01}
Benjamini, Y. and Yekutieli, D. (2001).
\newblock The control of the false discovery rate in multiple testing under
  dependency.
\newblock {\em The Annals of Statistics}, 29(4):1165--1188.

\bibitem[Berry and Berry, 2004]{Berry04}
Berry, S.~M. and Berry, D.~A. (2004).
\newblock Accounting for multiplicities in assessing drug safety: A three-level
  hierarchical mixture model.
\newblock {\em Biometrics}, 60(2):418--426.

\bibitem[Chen and Arias-Castro, 2017]{Chen17}
Chen, S. and Arias-Castro, E. (2017).
\newblock Sequential multiple testing.
\newblock \url{http://arxiv.org/abs/1705.10190}.

\bibitem[Chernoff, 1972]{Chernoff72}
Chernoff, H. (1972).
\newblock {\em Sequential Analysis and Optimal Design}.
\newblock Society for Industrial and Applied Mathematics, Philadelphia.

\bibitem[Cohen and Sackrowitz, 2005]{Cohen05}
Cohen, A. and Sackrowitz, H.~B. (2005).
\newblock Decision theory results for one-sided multiple comparison procedures.
\newblock {\em Annals of Statistics}, 33:126--144.

\bibitem[Dantzig, 1940]{Dantzig40}
Dantzig, G.~B. (1940).
\newblock On the non-existence of tests of {S}tudent's hypothesis having power
  functions independent of $\sigma$.
\newblock {\em The Annals of Mathematical Statistics}, 11:186--192.

\bibitem[Efron and Tibshirani, 2002]{Efron02}
Efron, B. and Tibshirani, R. (2002).
\newblock Empirical {B}ayes methods and false discovery rates for microarrays.
\newblock {\em Genetic Epidemiology}, 23(1):70--86.

\bibitem[Efron et~al., 2001]{Efron01}
Efron, B., Tibshirani, R., Storey, J.~D., and Tusher, V. (2001).
\newblock Empirical {B}ayes analysis of a microarray experiment.
\newblock {\em Journal of the American Statistical Association},
  96(456):1151--1160.

\bibitem[Espeland et~al., 2004]{Espeland04}
Espeland, M.~A., Rapp, S.~R., Shumaker, S.~A., Brunner, R., Manson, J.~E.,
  Sherwin, B.~B., Hsia, J., Margolis, K.~L., Hogan, P.~E., Wallace, R., et~al.
  (2004).
\newblock Conjugated equine estrogens and global cognitive function in
  postmenopausal women: {W}omenÕs {H}ealth {I}nitiative {M}emory {S}tudy.
\newblock {\em Journal of the American Medical Association},
  291(24):2959--2968.

\bibitem[Fischl et~al., 1987]{Fischl87}
Fischl, M.~A. et~al. (1987).
\newblock The efficiency of azidothymidine ({AZT}) in the treatment of patients
  with {AIDS} and {AIDS}-related complex.
\newblock {\em New England Journal of Medicine}, 317:185--191.

\bibitem[Genovese and Wasserman, 2002]{Genovese02}
Genovese, C. and Wasserman, L. (2002).
\newblock Operating characteristics and extensions of the false discovery rate
  procedure.
\newblock {\em Journal of the Royal Statistical Society: Series B (Statistical
  Methodology)}, 64(3):499--517.

\bibitem[Hoel et~al., 1971]{Hoel71}
Hoel, P.~G., Port, S.~C., and Stone, C.~J. (1971).
\newblock {\em Introduction to Statistical Theory}.
\newblock Houghton Mifflin Co., Boston, Mass.

\bibitem[Javanmard and Montanari, 2018]{Javanmard18}
Javanmard, A. and Montanari, A. (2018).
\newblock Online rules for control of false discovery rate and false discovery
  exceedance.
\newblock {\em The Annals of Statistics}, 46(2):526--554.

\bibitem[Jennison and Turnbull, 1997]{Jennison97}
Jennison, C. and Turnbull, B.~W. (1997).
\newblock Group sequential analysis incorporating covariate information.
\newblock {\em Journal of the American Statistical Association}, 92:1330--1341.

\bibitem[Jennison and Turnbull, 2000]{Jennison00}
Jennison, C. and Turnbull, B.~W. (2000).
\newblock {\em Group Sequential Methods with Applications to Clinical Trials}.
\newblock Chapman \& Hall/CRC, New York.

\bibitem[Lehmann and Romano, 2005]{Lehmann05}
Lehmann, E.~L. and Romano, J.~P. (2005).
\newblock {\em Testing Statistical Hypotheses}.
\newblock Springer, New York, third edition.

\bibitem[Mukhopadhyay and De~Silva, 2009]{Mukhopadhyay09}
Mukhopadhyay, N. and De~Silva, B. (2009).
\newblock {\em Sequential Methods and Their Applications}.
\newblock Chapman \& Hall/CRC.

\bibitem[M\"uller et~al., 2007]{Muller07}
M\"uller, P., Parmigiani, G., and Rice, K. (2007).
\newblock {FDR} and {B}ayesian multiple comparisons rules.
\newblock In Bernardo, J.~M., Bayarri, M.~J., Berger, J.~O., Dawid, A.~P.,
  Heckerman, D., Smith, A. F.~M., and West, M., editors, {\em {B}ayesian
  Statistics 8: Proceedings of the Eighth {V}alencia {I}nternational {M}eeting,
  June 2-6, 2006}, pages 349--370. Oxford University Press.

\bibitem[Newton et~al., 2004]{Newton04}
Newton, M.~A., Noueiry, A., Sarkar, D., and Ahlquist, P. (2004).
\newblock Detecting differential gene expression with a semiparametric
  hierarchical mixture method.
\newblock {\em Biostatistics}, 5(2):155--176.

\bibitem[O'Brien, 1984]{OBrien84}
O'Brien, P.~C. (1984).
\newblock Procedures for comparing samples with multiple endpoints.
\newblock {\em Biometrics}, 40:1079--1087.

\bibitem[Rossouw et~al., 2002]{Rossouw02}
Rossouw, J.~E., Anderson, G.~L., Prentice, R.~L., LaCroix, A.~Z., Kooperberg,
  C., Stefanick, M.~L., Jackson, R.~D., et~al. (2002).
\newblock Risks and benefits of estrogen plus progestin in healthy
  postmenopausal women: principal results from the {W}omen's {H}ealth
  {I}nitiative randomized controlled trial.
\newblock {\em Journal of the American Medical Association}, 288(3):321--333.

\bibitem[Sarkar, 1998]{Sarkar98}
Sarkar, S.~K. (1998).
\newblock Some probability inequalities for ordered $\rm {MTP}\sb 2$ random
  variables: a proof of the {S}imes' conjecture.
\newblock {\em The Annals of Statistics}, 26(2):494--504.

\bibitem[Shumaker et~al., 1998]{Shumaker98}
Shumaker, S.~A., Reboussin, B.~A., Espeland, M.~A., Rapp, S.~R., McBee, W.~L.,
  Dailey, M., Bowen, D., Terrell, T., and Jones, B.~N. (1998).
\newblock The {W}omenÕs {H}ealth {I}nitiative {M}emory {S}tudy ({WHIMS}): A
  trial of the effect of estrogen therapy in preventing and slowing the
  progression of dementia.
\newblock {\em Controlled Clinical Trials}, 19(6):604--621.

\bibitem[Siegmund, 1985]{Siegmund85}
Siegmund, D. (1985).
\newblock {\em Sequential Analysis: Tests and Confidence Intervals}.
\newblock Springer-Verlag, New York.

\bibitem[Siegmund and Yakir, 2008]{Siegmund08}
Siegmund, D. and Yakir, B. (2008).
\newblock Detecting the emergence of a signal in a noisy image.
\newblock {\em Statistics and Its Inference}, 1:3--12.

\bibitem[Simes, 1986]{Simes86}
Simes, R.~J. (1986).
\newblock An improved {B}onferroni procedure for multiple tests of
  significance.
\newblock {\em Biometrika}, 73(3):751--754.

\bibitem[Sonesson, 2007]{Sonesson07}
Sonesson, C. (2007).
\newblock A {CUSUM} framework for detection of space--time disease clusters
  using scan statistics.
\newblock {\em Statistics in Medicine}, 26(26):4770--4789.

\bibitem[Stein, 1945]{Stein45}
Stein, C. (1945).
\newblock A two-sample test for a linear hypothesis whose power is independent
  of the variance.
\newblock {\em The Annals of Mathematical Statistics}, 16:243--258.

\bibitem[Storey, 2002]{Storey02}
Storey, J.~D. (2002).
\newblock A direct approach to false discovery rates.
\newblock {\em Journal of the Royal Statistical Society: Series B (Statistical
  Methodology)}, 64(3):479--498.

\bibitem[Storey et~al., 2004]{Storey04}
Storey, J.~D., Taylor, J.~E., and Siegmund, D. (2004).
\newblock Strong control, conservative point estimation and simultaneous
  conservative consistency of false discovery rates: a unified approach.
\newblock {\em Journal of the Royal Statistical Society: Series B (Statistical
  Methodology)}, 66(1):187--205.

\bibitem[Storey and Tibshirani, 2003]{Storey03}
Storey, J.~D. and Tibshirani, R. (2003).
\newblock Statistical significance for genomewide studies.
\newblock {\em Proceedings of the National Academy of Sciences},
  100(16):9440--9445.

\bibitem[Wald, 1947]{Wald47}
Wald, A. (1947).
\newblock {\em Sequential Analysis}.
\newblock Wiley, New York.
\newblock Reprinted by Dover, 1973.

\bibitem[Woodall, 2006]{Woodall06}
Woodall, W.~H. (2006).
\newblock The use of control charts in health-care and public-health
  surveillance.
\newblock {\em Journal of Quality Technology}, 38(2):89--104.

\end{thebibliography}

\def\cprime{$'$}

\end{document}